\documentclass[12pt,letterpaper]{article}
\pdfoutput=1
\usepackage{jheppub}
\usepackage{amssymb}
\usepackage{color}
\usepackage{slashed}
\usepackage{amsmath}
\usepackage{amsfonts}
\usepackage{amssymb}
\usepackage{graphicx,subcaption}
\usepackage{caption}
\usepackage{subcaption}
\usepackage{float}
\usepackage{tikz}
\usetikzlibrary[arrows]
\usepackage{amsmath}
\usepackage{epsfig}
\usepackage{appendix}
\usepackage{listings}
\usepackage{xcolor}
\newcommand{\vev}[1]{{\left< {#1} \right>}}
\newcommand{\be}{\begin{equation}}
\newcommand{\ee}{\end{equation}}

\def\bI {\mathbb{I}}

\newcommand{\traza}[1]{{\mathrm{Tr}\,a^{#1}}}
\newcommand{\gYM}{g_{\scriptscriptstyle{\mathrm{YM}}}}
\newcommand{\tlambda}{\tilde \lambda}

\title{The planar limit of $\mathcal{N}=2$ superconformal field theories}

\author{Bartomeu Fiol,}
\author{Jairo Mart\'inez-Montoya}
\author{and Alan Rios Fukelman}

\affiliation{Departament de F{\'\i}sica Qu\`antica i Astrof\'isica i \\Institut de Ci{\`e}ncies del Cosmos, 
Universitat de Barcelona,
Mart{\'\i}\ i Franqu{\`e}s 1, 08028 Barcelona, Catalonia, Spain}

\emailAdd{bfiol@ub.edu}
\emailAdd{jmartinez@icc.ub.edu}
\emailAdd{ariosfukelman@icc.ub.edu}

\abstract{We obtain the perturbative expansion of the free energy on $S^4$ for four dimensional Lagrangian ${\cal N}=2$ superconformal field theories, to all orders in the 't Hooft coupling, in the planar limit. We do so by using supersymmetric localization, after rewriting the 1-loop factor as an effective action involving an infinite number of single and double trace terms. The answer we obtain is purely combinatorial, and involves a sum over tree graphs. We also apply these methods to the perturbative expansion of the free energy at finite $N$, and to the computation of the vacuum expectation value of the 1/2 BPS circular Wilson loop, which in the planar limit involves a sum over rooted tree graphs.}

\begin{document}

\maketitle

\section{Introduction}

Part of the theoretical appeal of supersymmetric gauge theories is that, for certain questions, they allow more analytical control than their non-supersymmetric counterparts. An outstanding example is supersymmetric localization, which allows to reduce the evaluation of certain quantities of 4d ${\cal N}=2$ super Yang-Mills (SYM) theories to matrix integrals \cite{Pestun:2007rz}. For instance, the partition function on $S^4$ is reduced to
\begin{equation}
		Z_{S^4} = \int da \, e^{-\frac{8\pi^2}{g_{\text{YM}}^2} \textnormal{Tr}(a^2)} \mathcal{Z}_{\text{1-loop}} \, \lvert \mathcal{Z}_{\text{inst}} \lvert^2
\label{zs4intro}		
\end{equation} 
where $\mathcal{Z}_{\text{1-loop}}$ is a factor that arises from a 1-loop computation, while $\mathcal{Z}_{\text{inst}}$ is the instanton contribution. Similarly, the expectation value of a 1/2 BPS circular Wilson loop $\vev{W_R}$ is also reduced to a matrix integral  \cite{Pestun:2007rz}.

The fact that four dimensional questions admit zero dimensional answers constitutes a dramatic simplification, but still leaves the formidable task of evaluating these matrix integrals. A first approach consists of restricting the integrals to a Cartan subalgebra of the Lie algebra. In a second approach \cite{Billo:2017glv, Billo:2018oog, Billo:2019fbi},  the integrals are over the full Lie algebra, and the 1-loop factor in (\ref{zs4intro}) is rewritten as an effective action.

For ${\cal N}=4$ super Yang Mills theories, both $\mathcal{Z}_{\text{1-loop}}=1$ and $\lvert \mathcal{Z}_{\text{inst}} \rvert^2=1$ in (\ref{zs4intro}) \cite{Pestun:2007rz}. The free energy can be easily computed \cite{Russo:2012ay}, but $\vev{W_R}$ is less trivial. Using the first approach mentioned above, the vev of the 1/2 BPS circular Wilson loop can be computed for different gauge groups $G$ and representations $R$, on a case by case basis \cite{Drukker:2000rr, Fiol:2013hna, Fiol:2014fla}. Recently, using the second approach, we derived a general expression for $\vev{W_R}$ valid for all gauge groups $G$ and representations $R$, thus unifying and extending all previous exact results \cite{Fiol:2018yuc}.

For generic ${\cal N}=2$ super Yang-Mills theories, the evaluation of (\ref{zs4intro}) or $\vev{W_R}$ is considerably much more complicated. Within the first approach, the free energy and $\vev{W_R}$ for various representations have been evaluated with a saddle point approximation \cite{Rey:2010ry, Russo:2012ay, Buchel:2013id, Russo:2013kea, Passerini:2011fe, Bourgine:2011ie, Fiol:2015mrp}. In this note we will apply the second approach to the study of $\mathcal{N}=2$ Lagrangian superconformal field theories (SCFTs) for arbitrary gauge groups at finite $N$, and for classical gauge groups in the planar limit. Ideally, we would like to write the quantities of interest in terms of color invariants of the gauge group and matter representations. This is vastly more complicated than in the ${\cal N}=4$ case considered in our previous work \cite{Fiol:2018yuc}, because the matrix model is interacting, and the identification of its perturbative expansion with the usual one in field theory - in terms of Feynman diagrams - is not immediate.

Let's outline our strategy and our results in some detail. Following ideas presented in \cite{Billo:2017glv, Billo:2018oog, Billo:2019fbi}, in Section 2 we rewrite the 1-loop factor $\mathcal{Z}_{\text{1-loop}}$ in (\ref{zs4intro}) as an effective action with an infinite number of single trace and double trace terms, where the traces are in the fundamental representation of the gauge group
\be 
\label{effactionintro}
\begin{split}
S_{int}^G =-\ln \mathcal{Z}_{\text{1-loop}}= \sum_{n=2}^\infty \frac{\zeta(2n-1)(-1)^n}{n}
\left[ (4-4^n)\alpha_G \traza{2n}+\beta_G \sum_{k=1}^{n-1}{2n \choose 2k} \traza{2(n-k)} \traza{2k} \right. \\
\left. + \gamma_G \sum_{k=1}^{n-2}{2n \choose 2k+1} \traza{2(n-k)-1} \traza{2k+1} \right],
\end{split}
\ee 
where $\alpha_G,\beta_G$ and $\gamma_G$ are constants that depend on the gauge group and the matter content of the conformal field theory. For the gauge group SU($N$), this effective action has been independently derived in \cite{Billo:2019fbi}, and can find applications beyond the ones presented in this work. 

Together with the kinetic term in (\ref{zs4intro}), the interaction terms in (\ref{effactionintro}) constitute a matrix model that is at the center of this work. Matrix models with single and double trace terms in the potential were discussed in the past \cite{Das:1989fq, Korchemsky:1992tt, AlvarezGaume:1992np, Klebanov:1994pv, Klebanov:1994kv}, in the context of two dimensional quantum gravity. In the planar limit, these models present different phases, depending on the relative strengths of couplings of the single and double trace terms.  For small coupling of the double trace term, the emerging geometry is that of a family of spheres connected by wormholes, created by the double trace terms \cite{Das:1989fq}. More specifically, the planar limit imposes that the full surface has genus zero, so the spheres connected by wormholes must form a tree graph, in the sense that no wormhole connects a sphere with itself, no two spheres are connected by more than a wormhole, and there is no closed loop of spheres \cite{AlvarezGaume:1992np, Klebanov:1994pv}. As the coupling of the double trace increases in the matrix model, the system develops new phases, including a branched polymer phase \cite{Das:1989fq}. 

The matrix model we encounter, with interaction terms (\ref{effactionintro}), bears some differences with the ones studied in the past \cite{Das:1989fq} and described above. First, the number of single and double trace terms in the effective action is now infinite. Moreover, the single trace terms in (\ref{effactionintro}) do not have the right scaling to contribute to the planar limit. Additionally, the coefficients of the single trace terms $\traza{2n}$ grow exponentially with $n$. On the other hand, the work \cite{Grassi:2014vwa}
does consider - in their Appendix B - a matrix model with an infinite number of double trace terms, and studies it in the planar limit with the technique of orthogonal polynomials.\footnote{We would like to thank Marcos  Mari\~no for pointing out this reference to us.}

Armed with this effective action (\ref{effactionintro}), we set out to evaluate various quantities of interest. The first one is the free energy of these SCFTs on $S^4$. As the integrals are Gaussian, they can be easily carried out. At finite $N$ what is left is the evaluation of color invariants in the fundamental representation. There are well-known techniques to help with the evaluation of these traces \cite{vanRitbergen:1998pn, Fiol:2018yuc}, but the expressions become more and more cumbersome as we go to higher orders in the perturbative expansion. Furthermore, the resulting expressions involve color invariants of the fundamental and adjoint representations, not of the matter representations of the various field theories. We then turn to the planar limit, and argue that only for theories with a finite fraction of matter in the fundamental representation - theories with $\beta_G\neq 0$ in (\ref{effactionintro}) - the planar free energy differs from the ${\cal N}=4$ result. For these theories, we manage to write the full perturbative expansion to all orders in the 't Hooft coupling $\lambda$,

\begin{multline}
F_0(\lambda)-F_0(\lambda)^{{\cal N}=4}=\sum_{n=2}^\infty \left(-\frac{\lambda}{16\pi^2}\right)^n \sum_{\substack{\text{compositions of n}\\ \text{not containing 1}}} (-2\beta_G)^m 
\frac{\zeta (2n_1-1)\dots \zeta(2n_m-1)}{n_1\dots n_m} \\
\sum _{k_1=1}^{n_1-1} {2n_1 \choose 2k_1} \dots
\sum _{k_m=1}^{n_m-1} {2n_m \choose 2k_m}  
\sum_{\substack{\text{unlabeled trees} \\ \text{with m edges}}} \frac{1}{|\text{Aut(T)}|}  {\cal V}_1 \dots {\cal V}_{m+1} 
\label{freeenergyintro}
\end{multline}
where $m$ is the number of elements of a given composition of $n$\footnote{A composition is a partition where the order of the elements matters; {\em e.g.} $2+3$ and $3+2$ are different compositions of 5.}, and ${\cal V}_i$ are combinatorial factors, to be defined below, attached to each of the $m+1$ vertices of the tree. As it turns out, (\ref{freeenergyintro}) involves a sum over tree graphs, thus making contact with the picture encountered in the context of two dimensional gravity. 

As a matter of fact, the particular values of the coefficients of the double trace terms in (\ref{effactionintro}), including the binomial coefficients, don't play any role in our argument, so we have effectively shown that the planar free energy of any matrix model with just double trace terms in the potential will involve the same sum over trees as (\ref{freeenergyintro}).

A basic question about this planar perturbative series (\ref{freeenergyintro}), is whether it has a non-zero radius of convergence $\lambda_c$, as expected on general grounds, and what is its precise value. Recall that in full-fledged quantum field theories, perturbative series are usually asymptotic, due to the combinatorial explosion of the number of Feynman diagrams. In the case at hand, the perturbative series are presumably divergent, but they are Borel summable \cite{Russo:2012kj, Honda:2016mvg}. On the other hand, there are generic arguments that in the planar limit, the drastic reduction of the number of diagrams implies that their number only grows powerlike with the number of loops, so the perturbative series has a finite radius of convergence \cite{Koplik:1977pf}. Finding the radius of convergence of (\ref{freeenergyintro}), and more generally, unveiling the phase structure of these theories in the planar limit, as the 't Hooft coupling  is varied, are important open questions.

In Section 3 we tackle the evaluation of the expectation value of the 1/2 BPS circular Wilson loop. Again, we start by computing the first terms in the perturbative expansion at finite $N$. Then we turn to the planar limit, and restrict ourselves to Wilson loops in the fundamental representation. We argue that $\vev{W_F}$ differs from the ${\cal N}=4$ one only for theories with a finite fraction of matter in the fundamental representation. We again manage to derive the perturbative expansion to all orders in $\lambda$; it now involves a sum over rooted trees.

In this work we have restricted ourselves to superconformal theories for concreteness. Looking towards the future, the techniques we have used can be also applied to non-conformal theories, massive or not. It will be interesting to determine whether any of the phase transition encountered for these theories \cite{Russo:2012ay, Buchel:2013id, Russo:2013kea} can be detected with our methods.

\section{The partition function of ${\cal N}=2$ superconformal Yang-Mills theories}
In this section we discuss the partition function of four dimensional Lagrangian ${\cal N}=2$ superconformal field theories on $S^4$. The seminal work \cite{Pestun:2007rz} showed that for Lagrangian ${\cal N}=2$ super Yang-Mills theories - not necessarily conformal - $Z_{S^4}$ can be reduced, thanks to supersymmetric localization, to a matrix integral. In this work we will consider  the perturbative expansion in the zero-instanton sector. We will follow the approach of \cite{Billo:2017glv, Billo:2018oog, Billo:2019fbi} and consider the integrals over the full Lie algebra. Furthermore, following also \cite{Billo:2017glv, Billo:2018oog, Billo:2019fbi} we rewrite the 1-loop factor of the integrand as an effective action. Our first result is a general expression for the complete effective action (see also \cite{ Billo:2019fbi} for the SU($N$) case). Armed with this result, we apply it first to obtain in a unified way the first terms of the partition function for classical Lie groups at finite $N$. We then switch to the planar limit, and obtain the planar free energy to all orders in the 't Hooft coupling.

We start by identifying the theories we will be studying. Lagrangian $\mathcal{N}=2$ super Yang-Mills theories with semi-simple gauge group $G$ and arbitrary matter hypermultiplets have been classified in \cite{Howe:1983wj,Koh1984, Bhardwaj2013}. The necessary and sufficient condition for  conformality of such theories is the vanishing $\beta$ function at 1-loop order, which translates into the following condition for the matter hypermultiplets

	\begin{equation}
		I_2(\textnormal{adj}) = \sum_{\mathcal{M}}  n_\mathcal{M} I_2(\mathcal{M}) 
	\label{def:beta0}
	\end{equation}
where $n_{\mathcal{M}}$ is the number of matter multiplets and the index of the representation $I_2(\mathcal{M})$ is defined in (\ref{def:conventions}). In this article we will be mainly interested in the classical groups for which (\ref{def:beta0}) reads
\begin{equation}\label{def:matter_group}
	\begin{split}
		\textnormal{SU}(N): \, &\qquad 2N = 2N n_{\mathrm{adj}} + n_F + (N+2)n_{\mathrm{sym}} + (N-2)n_{\mathrm{asym}} \\
		\textnormal{SO}(2N): \, &\qquad 2N-2 = (2N-2)n_{\mathrm{adj}} + n_{\mathrm{v}} \\
		\textnormal{SO}(2N+1): \, &\qquad 2N-1 = (2N-1) n_{\mathrm{adj}} + n_{\mathrm{v}} \\
		\textnormal{Sp(2N)}: \, &\qquad 2N+2 = (2N+2)n_{\mathrm{adj}} +  n_{\mathrm{v}} + (2N-2) n_{\mathrm{asym}},
	\end{split}
\end{equation}

\subsection{The 1-loop factor as an effective action}
As shown in \cite{Pestun:2007rz}, supersymmetric localization reduces the partition function of ${\cal N}=2$ SYM theories on $S^4$, to a matrix integral of the form 
\begin{equation}
		Z_{S^4} = \int da \, e^{-\frac{8\pi^2}{g_{\text{YM}}^2} \textnormal{Tr}(a^2)} \mathcal{Z}_{\text{1-loop}} \, \lvert \mathcal{Z}_{\text{inst}} \lvert^2
		\label{def:partition_function_squashing}
\end{equation} 
In (\ref{def:partition_function_squashing}), $da$ denotes a flat integration measure, over all the matrix entries,  $\mathcal{Z}_{\text{1-loop}}$ is a factor that arises from a 1-loop computation, while $|\mathcal{Z}_{\text{inst}}|^2$ is the instanton contribution. From now on we will restrict ourselves to the zero-instanton sector, for which $\lvert \mathcal{Z}_{\text{inst}} \rvert^2=1$. 

The $\mathcal{Z}_{\text{1-loop}}$ factor contains all the information of the choice of $G$ and matter, and it's given by products over the weights of the adjoint and of the matter representations
\begin{equation}
		\mathcal{Z}_{\text{1-loop}} = \prod_\alpha H(i \alpha \cdot \hat{a}) \prod_{R} \prod_{\omega_R} H(i \omega_R \cdot \hat{a})^{-n_R} 
	\label{def:1_loop}
	\end{equation}
where $H(x)$ is the Barnes function whose expansion is given by 
	\begin{equation}
		\ln H(x) = - (1+\gamma) x^2 - \sum_{n=2}^\infty \zeta(2n-1) \frac{x^{2n}}{n}
		\label{def:barnes_expansion}
	\end{equation}

Following \cite{Billo:2017glv, Billo:2018oog, Billo:2019fbi}, our strategy will be to rewrite the 1-loop partition function $\mathcal{Z}_{\text{1-loop}}$ in (\ref{def:partition_function_squashing}) as an effective action 
\begin{equation}
		S_{int}^G \equiv -\ln \mathcal{Z}_{\text{1-loop}} = S_2(a) + S_3(a) + \cdots
	\end{equation}
where each $S_i$ term corresponds to a value of the series expansion (\ref{def:barnes_expansion}), so effectively any computation is reduced to evaluation of correlators in the Gaussian theory. Due to the vanishing of the 1-loop $\beta$ function (\ref{def:beta0}) the sum of quadratic terms in (\ref{def:barnes_expansion}) cancel among themselves, so the effective action starts at order $g_{\text{YM}}^4$. The key step is that in (\ref{def:1_loop}), after carrying out the multiplications by the weights of the different representations, we get sums of products of eigenvalues of the matrix $a$. These products can be rewritten as products of traces of $a$ in the fundamental representation of $G$. Since the weights involved in (\ref{def:1_loop}) have one or two non-zero entries, these products translate into single trace and double trace operators respectively. For instance,
\be  
\sum_{u,v=1}^n (a_u+a_v)^{2n}= \sum_{u,v=1}^n \sum_{k=0}^{2n} {2n \choose k} a_u^{2n-k} a_v^k = \sum_{k=0}^{2n} {2n \choose k} \traza{2n-k} \traza{k}
\ee
Going through this procedure for all possible matter representations (\ref{def:matter_group}) is a straightforward but tedious exercise, so we leave the explicit form for  Appendix A. Here we present the general result, written in a unified manner for an arbitrary group $G$ as,
\be \label{def:general_action}
\begin{split}
S_{int}^G =\sum_{n=2}^\infty \frac{\zeta(2n-1)(-1)^n}{n}
\left[ (4-4^n)\alpha_G \traza{2n}+\beta_G \sum_{k=1}^{n-1}{2n \choose 2k} \traza{2(n-k)} \traza{2k} \right. \\
\left. + \gamma_G \sum_{k=1}^{n-2}{2n \choose 2k+1} \traza{2(n-k)-1} \traza{2k+1} \right],
\end{split}
\ee 
with coefficients $\alpha_G$, $\beta_G$, $\gamma_G$ that depend on the gauge group and matter content, see table (\ref{res:coefficients}). Eventually, we will be interested in taking the large $N$ limit of various quantities. Since for ${\cal N}=2$ SCFTs, $n_F$ scales with $N$, we have used the condition of the vanishing of the 1-loop $\beta$ function, eq. (\ref{def:beta0}), to eliminate $n_F$ in the previous formulas; in this way, all these coefficients are of order one, independent of $N$. Furthermore, they vanish for ${\cal N}=4$, as they should. The coefficient $\beta_G$ is essentially what was called $\nu$ in \cite{Fiol:2015mrp}, the fraction of matter in the fundamental representation.

\begin{table}[h!]
\centering
\begin{tabular}{|l|c|c|c|}
\hline
~ & $\alpha_G$ & $\beta_G$ & $\gamma_G$ \\
\hline
SU($N$) & $\frac{n_{\mathrm{sym}}-n_{\mathrm{asym}}}{2}$ & $1-n_{\mathrm{adj}}-\frac{n_{\mathrm{sym}}+n_{\mathrm{asym}}}{2}$ & $n_{\mathrm{adj}}-1-\frac{n_{\mathrm{sym}}+n_{\mathrm{asym}}}{2}$ \\
\hline
SO($2N$) & $1-n_{\mathrm{adj}}$ & $2(1-n_{\mathrm{adj}})$ & 0 \\
\hline
SO($2N+1$) & $1-n_{\mathrm{adj}}$ & $2(1-n_{\mathrm{adj}})$ & 0 \\
\hline
Sp($2N$) & $-1+n_{\mathrm{adj}}-n_{\mathrm{asym}}$ & $2(1-n_{\mathrm{adj}}-n_{\mathrm{asym}})$ & 0 \\
\hline
\end{tabular}
\caption{Value of the coefficients in (\ref{def:general_action}) for the different Lie groups.}
\label{res:coefficients}
\end{table}

We conclude that for any Lagrangian ${\cal N}=2$ SCFT, the 1-loop factor in (\ref{def:partition_function_squashing}) can be expressed as the exponential of an action that includes infinitely many single and double trace terms. For SU($N$), this all-order effective action was also recently derived in  \cite{Billo:2019fbi}.  As mentioned in the introduction, matrix models with single and double trace terms in the action already appeared in the study of two-dimensional quantum gravity \cite{Das:1989fq, Korchemsky:1992tt, AlvarezGaume:1992np, Klebanov:1994kv}, where these double traces were interpreted as wormholes connecting spheres. In the planar limit, for small enough coupling of the double trace term, the relevant surfaces were trees of spheres connected by these wormholes; we will see the reappearance of tree graphs in the planar limits of the free energy  - Section (\ref{secplanarf}) - and of the expectation value of the Wilson loop, Section (\ref{secplanarwilson}).

It is important to appreciate a difference between the matrix model that results from the effective interacting action (\ref{def:general_action}) and the matrix models just mentioned. Terms in the action of the matrix model that contribute of the large $N$ limit are - after perhaps a rescaling of the matrix - of the form
\be 
S = N^2 \, W ({\cal O})
\ee
where $W({\cal O})$ is a function that has no explicit $N$ dependence and ${\cal O}$ are normalized trace operators, {\em e.g.} ${\cal O}=\frac{1}{N} \traza{k}$.  Schematically, for a theory with kinetic term, single and double trace term interactions,
\be 
S = N^2 \left( \frac{1}{N} \traza{2} +\frac{1}{N} \traza{k} +\frac{1}{N^2} \traza{m} \, \traza{n} \right)
\label{properscaling}
\ee
The kinetic term in (\ref{def:partition_function_squashing}) is already of this form, since $\lambda = g_{\text{YM}}^2 N$. However, the single trace terms in (\ref{def:general_action}) do not have the proper scaling (\ref{properscaling}) to contribute to the planar limit. On the other hand, the double trace terms in (\ref{def:general_action}) do have the right scaling, and can contribute to the planar limit.

\subsection{Partition function and color invariants at finite $N$}
In this section we will compute the first terms of the zero-instanton sector of the normalized version of the partition function (\ref{def:partition_function_squashing}), using the explicit form of $S_{int}^G$ (\ref{def:general_action}), \emph{i.e.} 
	\begin{equation}
		Z_{S^4}=\frac{\vev{e^{-\sum_{i=2}^\infty S_i(a)}}_0}{\vev{\bI}_0}
		\label{eq:partition_expansion}
	\end{equation}
where the subscript $0$ corresponds to the Gaussian matrix model over the full Lie algebra. The denominator is the partition function for the case with $\mathcal{Z}_{\text{1-loop}}=1$, namely, the ${\cal N}=4$ SYM theory. By not restricting the integrals to the Cartan subalgebra \cite{Billo:2017glv,Billo:2018oog}, the Vandermonde determinant is not generated and the matrix integrals reduce to Gaussian ones that can be carried out by applying Wick's theorem. As discussed in \cite{Fiol:2018yuc}, this approach has the advantage that it allows to deal with different gauge groups and matter content in a unified fashion. 

We will actually compute the first terms of the free energy $F(\lambda,N)$, or to be more precise, due to the denominator in (\ref{eq:partition_expansion}), $F(\lambda,N)-F(\lambda,N)^{{\cal N}=4}$.  From a field theory perspective, it receives contributions only from connected Feynman diagrams.  At a given order in $g_{\text{YM}}$ the relevant interaction terms can be read off directly from (\ref{def:general_action}). As mentioned above, for superconformal field theories, the effective action starts at order $g^4_{\text{YM}}$, so the first cancellation when considering the logarithm of the partition function takes place at order $g^8_{\text{YM}}$. Up to this order,
\be
F(\lambda,N)-F(\lambda,N)^{{\cal N}=4}	 = -\vev{S_2(a)} - \vev{S_3(a)} - \vev{S_4(a)}+ \frac{1}{2}\left(\vev{S_2(a)^2}-\vev{S_2(a)}^2 \right) + \mathcal{O}(g^{10}_{\text{YM}})
\label{eq:logzfirst}
\ee
The computation factorizes into a trivial evaluation of Gaussian correlators, and the evaluation of color traces. The first part just amounts to applying Wick's theorem with the following two-point function,
\be 
\langle a_{a} a_{b} \rangle_0 = \frac{g^2_{\text{YM}}}{8\pi^2}\delta_{ab} 
\ee
The second one consists of evaluating traces of the Lie algebra generators, and it can be carried out using the techniques described in \cite{vanRitbergen:1998pn}. Our conventions are
	\begin{equation}
	 [T^{a}_R , T^{b}_R] = i f^{abc}T^c_R \, \qquad
			\textnormal{Tr}(T^a_R T^b_R) = I_2(R) \delta^{ab} \, \qquad      (T^a_R T^a_R)_{ij}=C_2(R) \delta_{ij}  
		\label{def:conventions}
	\end{equation}
with $a,b=1,\dots,N_A$, $N_R$ is the dimension of the representation $R$, and $A$ denotes the adjoint representation. We further define fully symmetrized traces
\be
d_R^{a_1\dots a_n}=\frac{1}{n!} \hbox{ Tr }\sum_{\sigma \in S_n} T_R^{a_{\sigma(1)}}\dots T_R^{a_{\sigma (n)}}
\ee  
To the order considered here, the relevant correlators (recall that all traces are in the fundamental representation) are
\be
\vev{\traza{2n}}=(2n-1)!! d_F^{b_1b_1\dots b_n b_n}
\ee
\be
\vev{\traza{2} \traza{2n}}=I_2(F)(d_A+2n)\vev{\traza{2n}}
\ee
\be
\vev{\traza{3} \traza{3}}=6 d_F^{abc} d_F^{abc}
\ee
\be
\vev{\traza{3} \traza{5}}=(60C_F - 15C_A)d_F^{abc}d_F^{abc} ,
\ee
and plugging them in (\ref{eq:logzfirst}) we obtain

\be
\begin{split}
F-F^{{\cal N}=4} = & -3\zeta(3)\frac{{\gYM}^4}{(8\pi^2)^2} \left[ \alpha_G(C_A - 6C_F) + \beta_G I_2(F)(2+N_A) \right]I_2(F)N_A \\
& +\zeta(5) \frac{{\gYM}^6}{(8\pi^2)^3} \left[-300\alpha_G d_{F}^{aabbcc}+30 \beta_G I_2(F) (N_A+4) d_F^{aabb}+40\alpha_G d_{F}^{abc}d_F^{abc}\right] \\
& -7 \zeta(7)\frac{g_{\text{YM}}^8}{(8\pi^2)^4} \left[-945 \alpha_G d_F^{aabbccdd} +30\beta_G I_2(F)(N_A+6) d_F^{aabbcc}+60\gamma_G (4C_F-C_A) d_F^{abc}d_F^{abc}\right. \\
& \left. +5 \beta_G \left(\frac{9}{2} d_F^{aabb}d_F^{ccdd} +I_2(F)^2 N_A (6C_F-C_A)^2 +12 d_F^{abcd}d_F^{abcd}\right) \right] \\
& + 18 \zeta(3)^2 \frac{g_{\text{YM}}^8}{(8\pi^2)^4} \Bigl[ \alpha_G^2 \left( 2I_2(F) N_A (6C_F - C_A)^2 +24 d_F^{abcd}d_F^{abcd} \right) \Bigr. \\
& \Bigl. -4 \alpha_G \beta_G I_2(F)^3 N_A(N_A+3)(6C_F-C_A) +2 \beta_G^2 I_2(F)^4 N_A (N_A+2)(N_A+3) \Bigr] +{\cal O}(g_{\text{YM}}^{10})
\end{split}
\label{res:logZ}
\ee

If needed, the color invariants that appear in the expression above can be rewritten in terms of lower order color invariants, as discussed in \cite{vanRitbergen:1998pn, Fiol:2018yuc}.  At order $g^{2n}_{\text{YM}}$ the possible products of values of the $\zeta$ function that can appear are $\zeta(2n_1-1)\dots \zeta(2n_m-1)$, where $\{n_1,\dots,n_m\}$ is a partition of $n$ not containing 1. The number of such partitions is $p(n)-p(n-1)$, where $p(n)$ is the number of partitions of $n$ \footnote{The generating function of the number of partitions of $n$ not containing 1 is $\prod_{k=2}^\infty \frac{1}{1-x^k}=(1-x)\prod_{k=1}^\infty \frac{1}{1-x^k} =(1-x) \sum_n p(n)x^n =\sum_n \left(p(n)-p(n-1)\right)x^n$.} .

The drawback of the approach we have pursued to carry out the integrals is that (\ref{res:logZ}) involves color invariants in the fundamental and adjoint representations, and not color invariants of the original matter representations of the SCFT. Therefore, it is not straightforward to match the different terms we encounter with the perturbative series in field theory. At low orders in the pertubative expansion, we can undo this, by rewriting the coefficient in terms of the original invariants. For instance, for the coefficient at order $g_{\text{YM}}^4$,
\be 
\alpha (C_A-6 C_F)+\beta I_2(F) (N_A+2)= C_A^2-\sum_R n_R C_R I_2(R)
\ee

If we now wish to study the large $N$ expansion of (\ref{res:logZ}), it is straightforward to evaluate this expression by fixing $G$, the matter content, computing the corresponding group factors and finally evaluating the large $N$ limit. For ${\cal N}=2$ SQCD, \emph{i.e.} taking $G=\text{SU}(N)$ with $n_F = 2N$, the corresponding group invariants are given by
	\begin{equation}\label{res:color factors SQCD}
			C_F = \frac{N^2-1}{2N} \,, \qquad N_A = N^2 - 1 \,, \qquad I_2(F) = \frac{1}{2} ,
	\end{equation}
In addition, from table (\ref{res:coefficients}) we see that this case corresponds to $\alpha_G = 0$, $\beta_G=1$ and $\gamma_G=-1$. All in all, taking the large $N$ limit and neglecting subleading terms
\be
\label{res:largeN partition function}
\begin{split}
F_0(\lambda)-F_0(\lambda)^{{\cal N}=4} =\,&-\frac{3 \zeta(3) }{256\pi^4} \lambda^2 + \frac{5\zeta(5)}{1024\pi^6} \lambda^3
  +  \frac{9\zeta(3)^2 -35\zeta(7)}{16384\pi^8} \lambda^4 +{\cal O}(\lambda^5)
\end{split}
\ee
where $F_0(\lambda)$ is the coefficient of $N^2$ in the $1/N$ expansion of the free energy.
 
\subsection{Free energy at large $N$}
\label{secplanarf}
We turn now our attention to the large $N$ limit of the free energy on $S^4$, $F(\lambda,N) =\ln Z_{S^4}$. The free energy admits a large $N$ expansion, $F(\lambda,N)=F_0(\lambda)N^2+\dots$, and our goal is to determine $F_0(\lambda)$. We will argue that $F_0(\lambda)$ differs from the ${\cal N}=4$ result only for theories with a finite fraction of matter in the fundamental representation, {\it i.e.} theories with $\beta_G\neq 0$ in (\ref{def:general_action}). In general,
\be 
F(\lambda,N)= \ln Z_{S^4} = \sum_{m=1}^\infty \frac{(-1)^{m+1}}{m} \left(\sum_{k=1}^\infty \frac{1}{k!} \vev{(-S^G_{int})^k}\right)^m
\label{freeseries}
\ee
In the previous expansion, $\vev{(S^G_{int})^k}$ involves disconnected $2k$-point functions whose $1/N$ expansion has a leading $N^{2k}$ term. On the other hand, the leading term in $F(\lambda,N)$ scales like $N^2$, so there are massive cancellations in (\ref{freeseries}). For actions with just single trace interactions, only planar connected diagrams contribute to $F_0(\lambda)$. The action (\ref{def:general_action}) has however double trace terms, and we need to fully identify the $N^2$ terms that survive the cancellations in (\ref{freeseries}). 

These contributions can be written as products of connected correlators, and as it turns out, the characterization of which products of connected correlators contribute to $F_0(\lambda)$ has a natural answer in terms of graph theory: for any product of connected correlators we introduce an associated graph, and we will argue that a product of connected correlators contributes to $F_0(\lambda)$ if and only if its associated graph is a tree. The perturbative expansion we find for $F_0(\lambda)$ of these theories is thus given by a sum over all tree graphs.

Once we accomplish the task of characterizing the contribution of any correlator to the planar free energy, we take advantage of the results of \cite{tutte, Gopakumar:2012ny} for the planar limit of connected correlators, and write the full perturbative expansion of the planar free energy of these theories.

As a starting point, notice that the different terms in $S_{int}^G$, eq. (\ref{def:general_action}), have vevs with different large $N$ scaling. The single trace terms  have vevs that scale like $N$. The vev of double trace operators with even powers factorizes in the large $N$ and it scales like $N^2$. On the other hand, the vev of a double trace of operators with odd powers does not have a disconnected contribution, and its leading term scales like $N^0$. This already suggests that the large $N$ behavior of the free energy depends qualitatively of having $\beta_G \neq 0$ or not; this qualitative difference was already encountered with the saddle point approximation. 

For ${\cal N}=4$ SYM, all coefficients in (\ref{def:general_action}) vanish, since the 1-loop factor is exactly one. The unnormalized partition function is then trivially given by Gaussian integrals, and the planar free energy takes the following form \cite{Russo:2012ay}
\be \label{eq:free_energy}
F_0(\lambda)^{{\cal N}=4}=\frac{1}{2}\ln \lambda
\ee
Let us discuss now some genuinely ${\cal N}=2$ SCFTs. When $\beta_G=0$, there are no terms in $S_{int}^G$ scaling like $N^2$, so $F_0(\lambda)=F_0(\lambda)^{{\cal N}=4}$. The last and most interesting case is that of theories with $\beta_G\neq 0$ in (\ref{def:general_action}), that is, with a finite fraction of matter in the fundamental representation. Theories with $\beta_G\neq 0$ can have $\alpha_G$ and $\gamma_G\neq 0$. We argue below that the $\alpha_G, \gamma_G$ parts of $S_{int}^G$ in eq. (\ref{def:general_action}) do not contribute to the free energy in the planar limit.

%\textcolor{red}{The case $\alpha_G=\beta_G=0$ is very confusing. The double trace of odd powers gives conflicting results: the 2-point of \cite{Gopakumar:2012ny} matches the graphical computation, but it disagrees with the large $N$ of the exact result above. The reason seems to be that the exact result above assumes that the generators are traceless. If they are not, there are additional terms proportional to $\mathrm{Tr}\, T \mathrm{Tr}\, T$. }

Any  disconnected correlator can be written as a sum of products of connected correlators
\begin{equation}
\vev{\traza{k_1} \dots \traza{k_n}} = \sum \vev{\traza{k_1} \dots \traza{k_{r_1}}}_c \dots \vev{\traza{k_{r_s}} \dots \traza{k_n}}_c
\label{eq:largeN_fact}
\end{equation}
Terms in the previous sum that grow faster than $N^2$ are too disconnected and cancel out when taking the logarithm. Terms that scale slower than $N^2$ do not contribute to the planar limit of the free energy. To characterize the terms in (\ref{eq:largeN_fact}) that scale precisely like $N^2$, let's recall that at large $N$, planar diagrams can be drawn on a sphere, so they scale like $N^2$. They are associated to connected correlators, and in the conventions of (\ref{properscaling}), the Feynman rule for a trace operator inserts an additional factor of $N$, so
\be
\vev{N\traza{k_1} \dots N\traza{k_n}}_c \sim N^2
\ee
or equivalently, the connected $n$-point function of trace operators scales as
\be
\vev{\traza{k_1} \dots \traza{k_n}}_c \sim N^{2-n}
\label{connectedcorr}
\ee
as long as there is an even number of odd $k_i$; if the number of odd $k_i$ is odd, the correlator vanishes.  According to (\ref{connectedcorr}), a term in the expansion (\ref{eq:largeN_fact}) that involves the product of $s$ connected correlators of $r_1,r_2,\dots r_s$ sizes, scales as
\be
\vev{\traza{k_1} \dots \traza{k_{r_1}}}_c \vev{\traza{k_1} \dots \traza{k_{r_2}}}_c \dots \vev{\traza{k_1} \dots \traza{k_{r_s}}}_c \sim N^{2s-(r_1+\dots +r_s)}
\ee
For this term to have the right scaling as $N^2$, the total number of operators in the disconnected correlator, $r_1+\dots +r_s$ must be even, call it $2m$, and furthermore $2s-2m=2$, so $s=m+1$. Therefore, for a disconnected $2m$-point function, the terms with the right $N^2$ scaling are products of precisely $m+1$ connected correlators. 
A slightly different version of the argument is the following: if we rewrite the double trace as
\be
\traza{2n-2k}\traza{2k}=\frac{1}{N^2} N \traza{2n-2k}\, N \traza{2k}
\ee
we observe that each double trace insertion comes with a $\frac{1}{N^2}$ factor. Then, the $N^2$ scaling comes from $s$ connected blobs, each scaling like $N^2$, joined by $m$ wormholes, each weighted by $\frac{1}{N^2}$
\be
(N^2)^s \left(\frac{1}{N^2}\right)^m = N^2 \Rightarrow s=m+1
\ee
Since we are partitioning $2m$ operators into $m+1$ correlators, the number of such products is given by the number of partitions of $2m$ into precisely $m+1$ parts, $p_{m+1}(2m)$. This can be shown to be the same as the number of partitions of $m-1$, $p(m-1)$ \footnote{ $p_{m+1}(2m)= [x^{2m}] x^{m+1}\prod_{i=1}^{m+1}\frac{1}{(1-x^i)}=[x^{m-1}] \prod_{i=1}^{m+1}\frac{1}{(1-x^i)}=  [x^{m-1}] \prod_{i=1}^{\infty}\frac{1}{(1-x^i)}=p(m-1)$. }. 

We have just argued that for a disconnected $2m$-point function, the terms that have the right large $N$ scaling to contribute to $F_0(\lambda)$ are products of $m+1$ connected correlators. But not all such terms do actually contribute to $F_0(\lambda)$, since they may not survive the cancellations that take place in the sum (\ref{freeseries}). If a term in $\vev{(S_{int}^G)^n}$ factorizes into pieces that appear in a product of $\vev{(S_{int}^G)^m}$ of lower orders, it will be cancelled. So the terms in $\vev{(S_{int}^G)^n}$ that contribute to $F_0(\lambda)$ are products of connected correlators, such that none of these correlators appears at lower orders. A succint way to describe this condition uses the language of graph theory. For this reason, we are going to associate a graph to any product of connected correlators. 

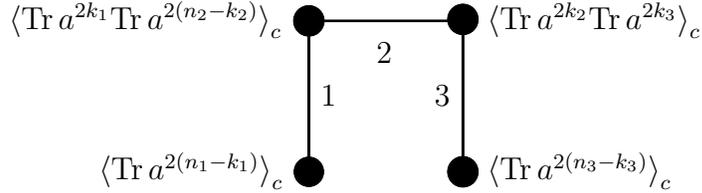
\begin{figure}[ht]
  \centering
     \includegraphics[scale=2]{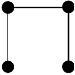}
    \put(3,5){${\langle \traza{2(n_3-k_3)} \rangle}_c$}
    \put(-60,33){$1$}
    \put(-17,33){$3$}
    \put(-39,49){$2$}
    \put(-144,5){${\langle \traza{2(n_1-k_1)} \rangle}_c$}
    \put(3,62){${\langle \traza{2k_2} \traza{2k_3} \rangle}_c $}
    \put(-178,62){${\langle \traza{2k_1} \traza{2(n_2-k_2)}  \rangle}_c$}
          \caption{How to map a product of connected correlators to a tree with labeled edges: For each connected correlator, introduce a vertex. If two vertices contain operators in the same double trace, join them by an edge. The edges are labeled by the double trace that the respective vertices have in common.}
    \label{corretotree}  
    \end{figure}

Consider a particular product of $m+1$ connected correlators of lengths $r_1,\dots, r_{m+1}$ such that $r_1+\dots+r_{m+1}=2m$. For each of them draw a vertex, so this is a graph with $m+1$ vertices. Then join two vertices by an edge if the correlators involve operators from the same double trace; there are then at most $m$ edges. See figure (\ref{corretotree}) for an example of this procedure.

\begin{figure}[ht]
\hspace{10ex}\begin{minipage}[c]{.33\textwidth}
  \begin{subfigure}{0.9\linewidth}
  	\vspace{-11ex}
    \centering
    \includegraphics[scale=1]{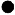}
    %\caption{$n=2$}
  \end{subfigure}
 
  \begin{subfigure}{0.9\linewidth}
  \vspace{-11ex}
   \centering
    \includegraphics[scale=1]{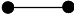}
    %\caption{$n=2$}
  \end{subfigure}
 
  \begin{subfigure}{0.9\linewidth}
  \vspace{-11ex}
   \centering
    \includegraphics[scale=1]{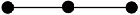}
   % \caption{$n=3$}
    \end{subfigure}
\hspace{\fill}
\end{minipage}%
\hspace{-12ex}\begin{minipage}[c]{.33\textwidth}
	\vspace{-11ex}	
  \begin{subfigure}{0.9\linewidth}
   \centering
    \includegraphics[scale=1]{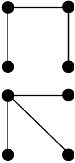}
  	%\caption{$n=4$}
  \end{subfigure}
\hspace{\fill}
\end{minipage}%
\hspace{-7ex}
\begin{minipage}[t]{.33\textwidth}
\begin{subfigure}{.2\linewidth}
\centering
    \includegraphics[scale=1]{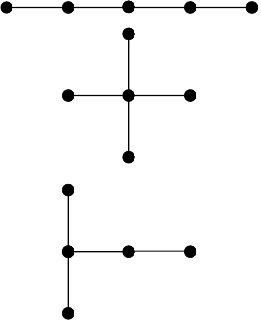}    
    %\put(-120,-15){(d) $n=5$}
  	  \end{subfigure}  
\end{minipage}
\caption{The list of trees up to 5 vertices.}
\label{sometrees}
\end{figure}

The condition on the correlators described above translates into the requirement that the graph is connected and has no loops; a connected graph with $m+1$ vertices and $m$ edges is a tree \cite{harary}. See figure (\ref{sometrees}) for the list of trees with up to five vertices.

\begin{figure}[ht]
	\centering
	\includegraphics[scale=1.3]{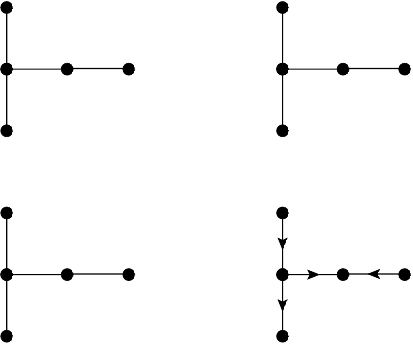}
	\put(-27,28){$4$}
	\put(-64.5,28){$3$}
	\put(-95,59){$1$}
	\put(-95,21){$2$}
	\put(-49.5,-10){$(d)$}
	%%%%%%%%%%%%%%%%%%%%%
	\put(-200.5,28){$4$}
	\put(-270,21){$2$}
	\put(-270,59){$1$}
	\put(-235,28){$3$}
	\put(-223.5,-10){$(c)$}
	%%%%%%%%%%%%%%%%%%%%%
	\put(-46,157){$4$}
	\put(-7,157){$5$}
	\put(-95,129.5){$3$}
	\put(-95,168){$2$}
	\put(-95,207){$1$}
	\put(-49.5,115){$(b)$}
	\put(-223.5,115){$(a)$}
	\caption{a) An unlabeled tree. b) A tree with labeled vertices. c) A tree with labeled edges. d) A directed tree with labeled edges.}
	\label{fig:botanica}
\end{figure}
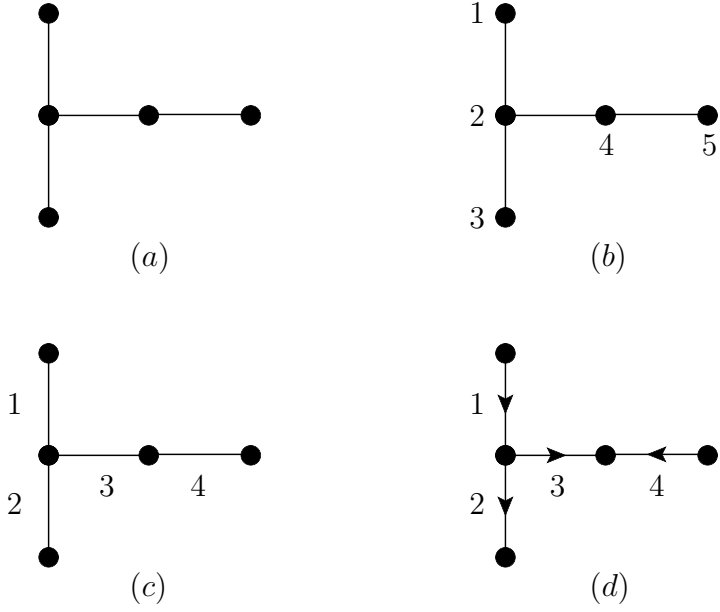

We can be more specific about the relevant types of trees. First, the edges are labeled by the double trace they represent. Furthermore, one has to distinguish two graphs coming from just swapping two operators in the same double trace. This can be taken into account by adding a direction (an arrow) to the edges, see figure (\ref{fig:botanica}). All in all, we have argued that the terms that contribute to $F_0(\lambda)$ from a disconnected $2m$-point function are in one-to-one correspondence with directed trees with $m+1$ vertices and labeled edges. 

Let's collect some basic results about the enumeration of trees. There is no known formula for the number of unlabeled trees with $n$ vertices. The sequence for the number of unlabeled trees with $n$ vertices has the following first few terms \cite{oeistrees}
\be 
1,1,1,2,3,6,11,23,\dots
\ee
A classical result by Cayley is that there are $(m+1)^{m-1}$ trees with labeled $m+1$ vertices \cite{cayleytrees}. Using this result, it is immediate to prove \cite{cameron} that for $m\geq 2$, there are $(m+1)^{m-2}$ trees with labeled edges. Finally, if every edge is oriented (with an arrow), there is an additional factor of 2 for each edge, so the number of oriented trees with $m+1$ vertices and labeled edges is $2^m (m+1)^{m-2}$ \cite{barequet}. See figure (\ref{fig:botanica}) for examples of these types of trees.

Let's go back to the expansion (\ref{freeseries}). Recall that the terms in the action (\ref{def:general_action}) don't have any power of $N$ in front of them. As discussed after eq. (\ref{properscaling}), that implies that double trace terms can contribute to the planar limit, but the single trace terms in (\ref{def:general_action}) can't. Let's further argue that only the double traces of even powers - the $\beta_G$ term in the action (\ref{def:general_action}) -  contribute to the planar limit. First, any non-zero correlator has an even number of odd powers, call it $2k$. In particular, no connected correlator can have just one odd power operator: it either has none, or at least two.  Therefore there are at most $k$ connected correlators with odd powers. The subgraph of connected correlators with odd powers has at most $k$ vertices and precisely $k$ edges, so it must contain loops. This implies that the full graph can't be a tree, and thus this product of connected correlators doesn't contribute to the large $N$ limit.

After arguing that only double traces of even powers contribute to the planar limit of the free energy, we restrict our attention to just those terms,
\be 
\begin{split}
Z_{S^4} =1+\sum_{m=1}^\infty \frac{(-1)^m \beta_G^m}{m!}\sum_{n_1,\dots,n_m=2}^\infty (-1)^{n_1+\dots+n_m}\frac{\zeta(2n_1-1)\dots \zeta(2n_m-1)}{n_1\dots n_m} \\
\sum_{k_1=1}^{n_1-1} {2n_1\choose 2k_1} \dots \sum_{k_m=1}^{n_m-1} {2n_m \choose 2k_m}  \vev{\traza{2(n_1-k_1)} \traza{2k_1}\dots \traza{2(n_m-k_m)} \traza{2k_m}}
\end{split}
\label{zbeforelog}
\ee
To proceed, we need the coefficients of connected correlators in the planar limit. These coefficients give the number of connected planar fatgraphs one can draw with the corresponding operators, and are thus integer numbers. For one-point functions \cite{Brezin:1977sv, DiFrancesco:2004qj}
\begin{equation}
\vev{\frac{1}{N} \traza{2k}} \rightarrow  C_k \left(\frac{\lambda}{16\pi^2}\right)^k 
\label{eq:largeN_catalan}
\end{equation}
where $C_k$ are the Catalan numbers. For connected $n$-point functions, the leading term at large $N$ is \cite{Gopakumar:2012ny} (see also \cite{tutte} for an earlier, purely combinatorial derivation ) 

	\begin{equation}
		\vev{\traza{2k_1} \traza{2k_2} \dots \traza{2k_n}}_c=\frac{(d-1)!}{(d-n+2)!} 
		\prod _{i=1}^n \frac{(2k_i)!}{(k_i-1)! k_i!} \left(\frac{\lambda}{16\pi^2}\right)^d N^{2-n} ,
		\label{gopakumarf}
	\end{equation}
where $d = \sum k_i$. Notice that (\ref{gopakumarf}) reduces to (\ref{eq:largeN_catalan}) when $n=1$. The results above were derived for the Hermitian matrix model, so in principle they apply to U($N$)/SU($N$) gauge theories. Nevertheless, since we are only concerned with planar diagrams, they apply also to SO($N$), Sp($N$) theories. For future use, let's give a name to the numerical coefficient in (\ref{gopakumarf}),
\be 
{\cal V}(k_1,\dots,k_n)= \frac{(d-1)!}{(d-n+2)!} 	\prod _{i=1}^n \frac{(2k_i)!}{(k_i-1)! k_i!}
\label{gpfactor}
\ee

%\begin{figure}[ht]
%  \centering
%  \hspace{5px}
%   \begin{subfigure}{0.3\textwidth}
%   \centering
%  \includegraphics[scale=1.5]{Trees/correlatorn_3.pdf}
%    \put(-13,-12){${\langle 2 \rangle}_c$}
%	\put(-62,-12){${\langle 1 2  \rangle}_c$}
%	\put(-103,-12){${\langle 1 \rangle}_c$}
%  \end{subfigure}
%  \begin{subfigure}{0.3\textwidth}
%  \centering
%     \includegraphics[scale=1.5]{Trees/correlatorn_4a.pdf}
%    \put(-12,-12){${\langle 3 \rangle}_c$}
%    \put(-56,-12){${\langle 1 \rangle}_c$}
%    \put(-15,58){${\langle 2 3 \rangle}_c $}
%    \put(-59,58){${\langle 1 2 \rangle}_c$}
%  \end{subfigure}
%  \hspace{-15px}
%  \begin{subfigure}{0.3\linewidth}
%  \centering
%    \includegraphics[scale=1.5]{Trees/correlatorn_4b.pdf}
%  	\put(-12,-12){${\langle 2 \rangle}_c$}
%    \put(-56,-12){${\langle 1 \rangle}_c$}
%    \put(-12,58){${\langle 3 \rangle}_c$}
%    \put(-62,58){${\langle 1 2 3 \rangle}_c $}
%  \end{subfigure}
%  \caption{More examples of the map between products of connected correlators and trees. }
% \end{figure} 

The contributions to $F_0(\lambda)$ at fixed order $\beta_G^m$ are then obtained as follows. At this order, there are $m$ pairs of traces, coming from $m$ double trace terms, $\traza{2n_1-2k_1}$, $\traza{2k_1},\dots$, $\traza{2n_m-2k_m}$,$\traza{2k_m}$. Draw all directed edge-labeled trees with $m$ edges. Assign $\traza{2n_i-2k_i}$ to the vertex at the start ({\it i.e}. origin of the arrow) of the i-th edge. Assign $\traza{2k_i}$ to the vertex at the end {\it i.e.} end of the arrow of the i-th edge. This procedure assigns to each of the $m+1$ vertices a number of traces equal to the degree of the vertex, {\it i.e.}  the number of edges connected to that vertex. For each vertex, consider now the connected correlator of all its trace operators and assign it its numerical factor ${\cal V}_i$, eq. (\ref{gpfactor}). Then,

\be 
\begin{split}
F_0(\lambda)-F_0(\lambda)^{{\cal N}=4} =\sum_{m=1}^\infty \frac{(-1)^m \beta_G^m}{m!}\sum_{n_1,\dots,n_m=2}^\infty \left( \frac{-\lambda}{16 \pi^2} \right)^{n_1+\dots+n_m}\frac{\zeta(2n_1-1)\dots \zeta(2n_m-1)}{n_1\dots n_m} \\
\sum_{k_1=1}^{n_1-1} {2n_1\choose 2k_1} \dots \sum_{k_m=1}^{n_m-1} {2n_m \choose 2k_m}  
\sum_{\substack{\text{directed trees}\\ \text{with labeled edges}}}
\prod_{i=1}^{m+1} {\cal V}_i
\end{split}
\label{planarf}
\ee
Let's illustrate this result by working out the lowest orders of (\ref{planarf}). Notice that $\beta_G$ counts the number of double trace terms, and it can be thought of as a wormhole counting parameter. When $m=1$, we have two-point functions, and we have to consider partitions of 2 into precisely 2 parts; the only possibility is 2=1+1, so the contribution comes from the product of two one-point functions.  Applying (\ref{eq:largeN_fact}) and (\ref{eq:largeN_catalan}) for the full series of (\ref{eq:partition_expansion}), we obtain

\begin{multline}
\left. F_0(\lambda) \right\vert_{\beta_G} = - \beta_G \sum_{n_1=2}^\infty \frac{\zeta(2n_1-1)}{n_1} \left( \frac{-\lambda}{16 \pi^2} \right)^{n_1}  \sum_{k_1=1}^{n_1-1} {2n_1 \choose 2k_1} C_{n_1-k_1}C_{k_1} \\
= - \beta_G \sum_{n=2}^\infty \frac{\zeta(2n-1)}{n} \left( \frac{-\lambda}{16 \pi^2} \right)^n C_n(C_{n+1}-2)
\label{linearbeta}
\end{multline}
as expected from the finite $N$ discussion we see that the leading $\beta_G$ term is an infinite series that captures all the terms with only one $\zeta$ thus generalizing the result (\ref{res:largeN partition function}). At order $\beta_G^2$, the terms that contribute to $F_0(\lambda)$ come from distributing the 4 operators into precisely 3 correlators. The only possible partition is $4=1+1+2$ , which corresponds to the only tree with 3 vertices in figure (\ref{sometrees}). There are 4 directed trees with labeled edges for this unlabeled tree, so the contributions are
	\begin{equation}
		\vev{\traza{2(n_1-k_1)}}_c \vev{\traza{2(n_2-k_2)}}_c \vev{\traza{2k_1} \traza{2k_2}}_c  
	\end{equation}
and the corresponding permutations coming from exchanging $k_i \leftrightarrow n_i-k_i$. While the product of correlators is not invariant under this exchange, after summing over $k_{1,2}$ in (\ref{planarf}), the answer is, so we can just take one of them and multiply by 4. The contribution to the planar free-energy is given by 

\begin{equation}								
\begin{split}
\left.	F_0(\lambda) \right\vert_{\beta_G^2} = \frac{\beta_G^2}{2!}  \sum_{n_1,n_2=2}^{\infty}\frac{\zeta(2n_1-1)\zeta(2n_2-1)}{n_1 n_2}\left( \frac{-\lambda}{16\pi^2} \right)^{n_1+n_2} \\
\sum_{k_1=1}^{n_1-1}{2n_1 \choose 2k_1} \sum_{k_2=1}^{n_2-1}{2n_2 \choose 2k_2} \frac{4}{k_1+k_2} \frac{(2k_1)!}{(k_1-1)! k_1!}
\frac{(2k_2)!}{(k_2-1)! k_2!} C_{n_1-k_1} C_{n_2-k_2}
\end{split}
\end{equation}
This expression recovers the term in (\ref{res:largeN partition function}) with a product of two values of $\zeta$, and provides all the subsequent terms of this form.

After these examples, let's simplify the sums in (\ref{planarf}). First, as we have seen in the example at order $\beta_G^2$, while the product ${\cal V}_1 \dots {\cal V}_{m+1}$, is not invariant under $k_i \leftrightarrow n_i-k_i$, after summing over all $k_i$ in (\ref{planarf}), the answer is the same for any of the choices of arrows of the tree, so one can just take any of the $2^m$ possible assignments, and replace the last sum by
\begin{equation}
2^m \sum_{\substack{\text{undirected trees}\\ \text{with labeled edges}}} {\cal V}_1 \dots {\cal V}_{m+1}
\end{equation}
where now the sum is over undirected trees (no arrows) with $m$ labeled edges. To further simplify this sum, note that for a given unlabeled tree T with $m\geq 2$ edges and with automorphism group $\text{Aut(T)}$, there are $\frac{m!}{|\text{Aut(T)}|}$ ways to label its edges \cite{cameron}. They correspond to different rearrengements of the indices $1,2,\dots,m$ in the traces placed in the $m+1$ correlators, so again, in general the values of the correlators are different. However, (\ref{planarf}) contains a sum over all $n_1,\dots, n_m$ so after this sum all such terms end up giving the same. The case $m=1$ has to be considered separately; the only tree with one edge, see figure (\ref{sometrees}), has $|\text{Aut(T)}|=2$ and there is just one way to label its edge. On the other hand, for the directed version, changing the direction of the arrow does not change the tree, so these two factors cancel each other. These considerations allow to further simplify the sum over trees to
\begin{equation}
2^m \sum_{\text{unlabeled trees}} \frac{m!}{|\text{Aut(T)}|} {\cal V}_1 \dots {\cal V}_{m+1}
\end{equation}
finally arriving at,
\begin{multline}
F_0(\lambda) -F_0(\lambda)^{{\cal N}=4}=  \sum_{m=1}^\infty (-2\beta_G)^m \sum _{n_1,\dots,n_m=2}^\infty \left( \frac{-\lambda}{16\pi^2} \right)^{n_1+\dots +n_m} \frac{\zeta (2n_1-1)\dots \zeta(2n_m-1)}{n_1\dots n_m} \\
\sum _{k_1=1}^{n_1-1} {2n_1 \choose 2k_1} \dots
\sum _{k_m=1}^{n_m-1} {2n_m \choose 2k_m}  
\sum_{\substack{\text{unlabeled trees} \\ \text{with m edges}}} \frac{1}{|\text{Aut(T)}|}  {\cal V}_1 \dots {\cal V}_{m+1}
\label{betaexpforf}
\end{multline}

A physically more relevant expression comes from grouping all terms with the same power of $\lambda$. To write it down, first recall that a composition of $n$ is a partition where order matters, so $3+2$ and $2+3$ are different compositions of 5. We will denote by $m$ the number of non-zero elements of a given composition. Then
\begin{multline}
F_0(\lambda)-F_0(\lambda)^{{\cal N}=4}=\sum_{n=2}^\infty \left(-\frac{\lambda}{16\pi^2}\right)^n \sum_{\substack{\text{compositions of n}\\ \text{not containing 1}}} (-2\beta_G)^m 
\frac{\zeta (2n_1-1)\dots \zeta(2n_m-1)}{n_1\dots n_m} \\
\sum _{k_1=1}^{n_1-1} {2n_1 \choose 2k_1} \dots
\sum _{k_m=1}^{n_m-1} {2n_m \choose 2k_m}  
\sum_{\substack{\text{unlabeled trees} \\ \text{with m edges}}} \frac{1}{|\text{Aut(T)}|}  {\cal V}_1 \dots {\cal V}_{m+1}
\label{finalfree}
\end{multline}
where the second sum is over compositions $\{n_1,\dots, n_m\}$ of $n$ that don't contain 1. The number of such compositions of $n$ is given by the Fibonacci number $F_{n-1}$ \cite{cayley}. Equation (\ref{finalfree}) is our result for the planar limit of the free energy of theories with $\beta_G\neq 0$, and the main result of this section. In Appendix B we write explicitly its first terms, up to 13th order.

Let's comment now on the convergence of the perturbative expansion (\ref{finalfree}). Typically, perturbative series in quantum field theory are asymptotic, due to the combinatorial explosion of Feynman diagrams. The perturbative series of the full free energy of these theories is presumably divergent, but it is Borel summable \cite{Russo:2012kj, Honda:2016mvg}. On the other hand, for generic quantum field theories, when we restrict to the planar limit, there is a drastic reduction in the number of Feynman diagrams that contribute in this limit, which now grows only powerlike.  As a consequence, the planar perturbative series has a finite radius of convergence \cite{Koplik:1977pf, Brezin:1977sv}. A pertinent question is then what is the radius of convergence of (\ref{finalfree}). 

We haven't been able to determine the radius of convergence of (\ref{finalfree}). Nevertheless, let us offer some comments on the convergence of the series that appear at every order in $\beta_G$ in (\ref{betaexpforf}). At every fixed order in $\beta_G$, the coefficient is a series in $\lambda$. At order $\beta_G$, it follows immediately from the quotient criterion that the series (\ref{linearbeta}) has radius of convergence $\lambda_c =\pi^2$. This is precisely the same value as the one found in \cite{Russo:2013kea} for the divergence of planar perturbation theory for ${\cal N}=4$ SYM generic observables (in this sense, the 1/2 BPS ${\cal N}=4$ Wilson loop turns out not to be a generic observable). We can sketch an argument proving that the series in $\lambda$ at generic, but fixed, order in $\beta_G$ in (\ref{betaexpforf}) have all the same radius of convergence. First, define ${\tilde {\cal V}}_i$ as the prefactor of ${\cal V}_i$ that does not factorize,
\be
\tilde {\cal V}_i(k_1,\dots,k_n)=\frac{(d-1)!}{(d-n+2)!}
\ee
with $d=\sum_i k_i$. Then, the last line in (\ref{betaexpforf}) can be rewritten as
\begin{multline}
\frac{(2n_1)!}{(n_1-1)!^2} \dots \frac{(2n_m)!}{(n_m-1)!^2}
\sum _{k_1=1}^{n_1-1} {n_1-1 \choose k_1} {n_1-1 \choose k_1-1} \dots
\sum _{k_m=1}^{n_m-1} {n_m-1 \choose k_m}{n_m-1 \choose k_m-1}  \\
\sum_{\substack{\text{unlabeled trees} \\ \text{with m edges}}} \frac{\tilde {\cal V}_1 \dots \tilde{\cal V}_{m+1}}{|\text{Aut(T)}|}  
\end{multline}
The sum over trees in the equation above yields a rational function of the variables $n_i,k_i$ of degree $-m-3$, let's call it $Q_m(n_i,k_i)$  For large $n_i$ the $m$ sums over $k_i$ can be thought as a measure peaked around $k_i=n_i/2$, so we conjecture that when all $n$s are large, the effect of the sums is evaluating $Q_m$ with all $k_i$ taking the value $n_i/2$
\be
\sum_{k_1} \sum_{k_m} Q_m(n_1,\dots,n_m,k_1,\dots,k_m) \rightarrow Q_m(n_1,\dots,n_m, n_1/2,\dots n_m/2) \sum_{k_1} \sum_{k_m}
\ee
If this is true, assuming all the $n_i$ are large enough and applying the Stirling approximation, it follows that for every $m$ the series at order $\beta_G^m$ in (\ref{betaexpforf}) has finite radius of convergence $\lambda_c =\pi^2$. Even if this argument can be made precise, proving that at every fixed order in $\beta$, the corresponding series in  (\ref{betaexpforf}) has radius of convergence $\lambda_c=\pi^2$ doesn't prove that this is the radius of convergence of (\ref{finalfree}). Study of the convergence of (\ref{finalfree}) is under investigation.

Finally, let's point out that the number of trace insertions at every vertex in the tree graph is fixed. It is the degree of the vertex, {\em i.e.} the number of edges arriving at the vertex. This is due to the fact that single trace terms don't contribute to the planar limit. 

\section{The 1/2 BPS circular Wilson loop}	

Another milestone of supersymmetric localization is the possibility to compute the expectation value of certain class of protected operators. In the work \cite{Pestun:2007rz} it was proven that the expectation value of the 1/2 BPS circular Wilson loop
\be 
W_R=\frac{1}{N_R} \textnormal{Tr}_R \mathcal{P} \, \exp \oint_C (A_\mu dx^\mu +i \Phi ds)
\ee 
also reduces to a matrix model computation
\begin{equation}
		\langle W \rangle = \frac{1}{Z_{S^4}}\int da \textnormal{Tr} e^{-2\pi b a} e^{-\frac{8\pi^2}{g_{\text{YM}}^2} \textnormal{Tr}(a^2)} \mathcal{Z}_{\textnormal{1-loop}} |\mathcal{Z}_{\textnormal{inst}}|^2
		\label{def:WL_MM}
	\end{equation}
It has been understood more recently that the correlator of the stress-energy tensor and a $1/2$ BPS circular Wilson loop can also be determined by a matrix model computation. First, the two-point function of the stress-energy tensor and a straight 1/2 BPS Wilson line is determined by conformal invariance, up to a coefficient $h_W$ \cite{Kapustin:2005py}
\be
\frac{\vev{T^{00}(x)W}}{\vev{W}}=\frac{h_W}{|\vec x|^4}
\ee
This coefficient appears also in the two-point function of the stress-energy tensor and a circular Wilson loop \cite{Fiol:2012sg}. It was conjectured in \cite{Fiol:2015spa} that for ${\cal N}=2$ SCFTs
\be 
h_W = \frac{1}{12\pi^2}\partial_b \ln \vev{W_b}  \vert_{b=1}
\label{hwfromwb}
\ee
where the vev of the Wilson loop is computed in a squashed $S^4$ sphere of parameter $b$,
\begin{equation}
		\langle W_b \rangle = \frac{1}{Z_{S^4}}\int da \textnormal{Tr} e^{-2\pi b a} e^{-\frac{8\pi^2}{g_{\text{YM}}^2} \textnormal{Tr}(a^2)} \mathcal{Z}_{\textnormal{1-loop}} |\mathcal{Z}_{\textnormal{inst}}|^2
		\label{def:WL_MMb}
	\end{equation}
In principle, both $\mathcal{Z}_{\textnormal{1-loop}}$ and $|\mathcal{Z}_{\textnormal{ins}}|^2$ depend on the squashing parameter $b$, but this dependence starts at quadratic order in $(b-1)^2$ \cite{Fiol:2015spa}.  In practice, since the relation (\ref{hwfromwb}) is only sensitive to the linear dependence in $b-1$, in evaluating (\ref{hwfromwb}) it is valid to use the expressions for $\mathcal{Z}_{\textnormal{1-loop}}$ and $|\mathcal{Z}_{\textnormal{ins}}|^2$ of the ordinary $S^4$. Additional evidence for eq. (\ref{hwfromwb}) was provided in \cite{Gomez:2018usu} and it was finally proven in \cite{Bianchi:2019dlw}. It is also worth keeping in mind that for ${\cal N}=2$ theories it was conjectured in \cite{Lewkowycz:2013laa, Fiol:2015spa} and proven in \cite{Bianchi:2018zpb} that $B=3 h_W$, where $B$ is the Bremsstrahlung function \cite{Correa:2012at, Fiol:2013iaa}.

The perturbative computation of the vev of this Wilson loop operator in $\mathcal{N}=2$ was studied before \cite{Andree:2010na} by usual QFT techniques, in \cite{Gomez:2018usu} by using the heavy quark effective theory and in \cite{Passerini:2011fe} by matrix model techniques. All of this perturbative computations were done for the case of $G=\text{SU}(N)$ with $n_F=2N$, in addition going to higher orders in perturbation theory seems a daunting task in these formalisms. 

As in the case of the partition function, we will attack this problem with localization techniques and once again we will not restrict the integration to the Cartan subalgebra of $G$. This will allow us to obtain both the Wilson loop operator and the Bremsstrahlung function in an unified manner for any choice of $G$ obeying (\ref{def:beta0}) in terms of color invariants. In the large $N$ limit, we will be able to obtain an all order expression in $\lambda$,  similar to the one found for the free energy in the previous section.  

We will consider the generalized Wilson loop $W_b$, eq. (\ref{def:WL_MMb}), on an ordinary $S^4$ so we can apply (\ref{hwfromwb}) to obtain the one-point function of the stress-energy tensor in the presence of the Wilson loop. To obtain the expectation value of the Wilson loop on $S^4$, it is enough to set $b=1$.

\subsection{Wilson loops at finite $N$}
In this section we will proceed as in the case of the free energy. We will perform a perturbative calculation of the lowest orders of the vev of the 1/2 BPS circular Wilson loop operator and we will be able to cast the result for arbitrary gauge group $G$ at finite $N$. Setting from now on $|\mathcal{Z}_{\textnormal{inst}}|^2=1$, from (\ref{def:WL_MMb}) we have

	\begin{equation}
		\langle W_b \rangle = \frac{1}{Z_{S^4}} \sum_{l=0}^\infty \frac{(-2\pi b)^l}{l! N_R} \langle \mathrm{Tr}_R\,a^{l} e^{-S(a)} \rangle_0,
		\label{eq:wl_expansion}
	\end{equation}	
since we are interested in corrections coming from the matter content it is convenient to subtract the expectation value of the Wilson loop operator of the $\mathcal{N}=4$ theory. Up to order $g_{\text{YM}}^8$ we find
	\begin{multline}
		\langle W \rangle_{\mathcal{N}=2}-\langle W \rangle_{\mathcal{N}=4} =
-\frac{1}{N_R}\frac{(2\pi b)^2}{2!} \Bigl[ \langle \mathrm{Tr}_R\,a^{2} S_2(a) \rangle - \langle \mathrm{Tr}_R\,a^{2} \rangle \langle S_2(a) \rangle + \vev{\mathrm{Tr}_R\,a^{2} S_3(a)} - \vev{\mathrm{Tr}_R\,a^{2}} \vev{S_3(a)} \Bigr] \\
-\frac{1}{N_R}\frac{(2\pi b)^4}{4!} \Bigl[ \langle \mathrm{Tr}_R\,a^{4} S_2(a) \rangle - \langle \mathrm{Tr}_R\,a^{4} \rangle \langle S_2(a) \rangle  \Bigr]
+{\cal O}(g_{\text{YM}}^{10})
		\label{eq:o6}
	\end{multline}

From (\ref{def:general_action}) it's a straightforward calculation to obtain at finite $N$
\begin{multline}
	\langle W \rangle_{\mathcal{N}=2}-\langle W \rangle_{\mathcal{N}=4} = 
	\frac{3\zeta(3) b^2 g_{\text{YM}}^6}{(8\pi^2)^2} \frac{I_2(R)}{N_R} \Bigl[6\alpha_G d_F^{aabb}-\beta_G I_2(F)^2 N_A (N_A+2)\Bigr] \\
	+\frac{\zeta(5)b^2 g_{\text{YM}}^8}{(8\pi^2)^3} \frac{I_2(R)}{N_R} \Bigl[-450 \alpha_G d_F^{aabbcc}+45 \beta_G I_2(F) (N_A+4)d_F^{aabb}+60 \gamma_G d_F^{abc}d_F^{abc} \Bigr] \\
	+ \frac{\zeta(3)b^4 g_{\text{YM}}^8}{4(8\pi^2)^4 N_R} \Bigl[-6\alpha_G (3\frac{d_R^{aabb}d_F^{ccdd}}{N_A}+d_R^{abcd}d_F^{abcd})+3\beta_G (N_A+3)d_R^{aabb} \Bigr]+{\cal O}(g_{\text{YM}}^{10})
\label{res:gg6}	
\end{multline}
	
As a check, we can compare the order $g_{\text{YM}}^6$ general result with	the computations carried out in \cite{Andree:2010na, Gomez:2018usu} for the special case of SQCD, this setup is the same as the one considered in (\ref{res:color factors SQCD}), with this is straightforward to evaluate the order $g_{\text{YM}}^6$ term in (\ref{res:gg6}) for this choice
	\be
		\langle W \rangle_{\mathcal{N}=2} - \langle W \rangle_{\mathcal{N}=4}
	 = -b^2\frac{g^6_{\text{YM}}}{512 \pi^4}(3\zeta(3))\frac{(N^2-1)(N^2+1)}{N} ,
	\ee
that precisely matches the result presented in \cite{Andree:2010na} and further generalizes it to any gauge group $G$ while preserving the finite $N$ contributions.  

% For the remaining groups of (\ref{def:beta0}) we note that the corresponding values of $\alpha, \beta, \gamma$ are the same and it's possible to write the expansion (\ref{eq:o6}) in a unified manner as
% 	\begin{equation}
% 		\langle W \rangle_{\mathcal{N}=4}-
% 	\langle W \rangle_{\mathcal{N}=2} = -3 b^2\frac{{\gYM}^6}{(8\pi^2)^2} \alpha_G  \Big[ C_a - 6C_f + 2 I_2(f)(2+d_A)) I_2(f) C_2(\mathcal{R}\Big]
% 	\label{result:g6}
% 	\end{equation}
As in the case of the free energy, a drawback of the result (\ref{res:gg6}) is that it's written in terms of gauge invariants in the fundamental representation, and not of the matter representations of the theory. To fix this, we can look at the relevant diagrams in the quantum field theory computation, eq. (18) in \cite{Andree:2010na}, to find that at order $g_{\text{YM}}^6$ the color factor is 
\be
\frac{C_A^2}{2}-\sum_h n_h (C_h -\frac{C_A}{2})I_2(h)=C_A^2- \sum_h n_h C_h I_2(h)
\ee
where the sum is over the matter hypermultiplets. It can be easily checked that the color invariant and the coefficients in (\ref{res:gg6}) do reproduce this color factor.

As a second example, consider the SCFT whose gauge group is SU($N$) and has one rank$-2$ symmetric and one rank$-2$ antisymmetric hypermultiplet, with this matter content the degrees of freedom scale as $\frac{1}{2}N(N+1)+\frac{1}{2}N(N-1) \simeq N^2$ this is the same number as $\mathcal{N}=4$ $SU(N)\times U(1)$ gauge theory. This coincidence leads to a massive cancelation of Feynman diagrams from which it's expected \cite{Bourget:2018fhe,Fiol:2013iaa} that corrections from the expectation value of the $\mathcal{N}=4$ result scale as $1/N$, this is easy to check noting that this theory corresponds (\ref{res:coefficients}) to $\alpha_G=0$, $\beta_G=0$ and $\gamma_G=-2$, so (\ref{res:gg6}) becomes
	\begin{equation}
		\vev{W}_{\mathcal{N}=2} - \vev{W}_{\mathcal{N}=4} = - \frac{120 b^2 g^8_{\text{YM}} \zeta(5)}{N_R (8\pi^2)^3} d_f^{abc} d_f^{abc}I_2(R) \simeq - \frac{\lambda^4}{4 N^2} \frac{15 b^2 \zeta(5)}{(8\pi^2)^3}	
	\end{equation}
where we see that corrections scale as $1/N^2$. 

Finally let us briefly discuss the large $N$ limit and present the result corresponding to SQCD, from (\ref{eq:o6}) and (\ref{res:gg6}) we obtain
\begin{equation}
	\begin{split}
	\langle W \rangle_{\mathcal{N}=2} - \langle W \rangle_{\mathcal{N}=4} = -\frac{3b^2 \lambda^3}{512 \pi^4}\zeta(3)  - \lambda^4 \left(\frac{2\pi^2 b^4 \zeta(3) - 15b^2 \zeta(5)}{4096\pi^6} \right) + {\cal O}(\lambda^5).
	\end{split}
\end{equation}

\subsection{Wilson loops at large $N$}
\label{secplanarwilson}
We want to determine now the planar limit of the expectation value of the Wilson loop operator. As found in the previous section for the planar free energy, the answer depends markedly on whether the SCFT has a finite fraction of matter in the fundamental representation, $\beta_G\neq 0$ in (\ref{def:general_action}), or not. 

Since we want to take advantage of the result for the planar connected correlators (\ref{gopakumarf}), we will restrict ourselves to the case $R=F$, so the Wilson loop is taken in the fundamental representation. Recall that the effective action (\ref{def:general_action}) involves single trace terms, and double traces of even and of odd power operators. Note that when $\beta_G\neq 0$, the $\alpha_G$ and $\gamma_G$ terms in the action give subleading contributions, and can be neglected in the planar limit.

First of all, let's argue that $\vev{W_b}$ scales like $N^0$ in the planar limit. Expanding the exponential of the Wilson loop insertion
\be
\langle W_b \rangle =  \sum_{l=0}^\infty \frac{(4\pi^2 b^2)^l}{(2l)!} \frac{\langle \frac{1}{N} \traza{2l} e^{-S}\rangle}{\vev{e^{-S}}}
\ee
Now, the expression inside the sum is the one-point function of $\frac{1}{N}\traza{2l}$ in an interacting matrix model, that scales like $N^0$. Following the same logic as in the previous section for the free energy, we aim to write this as a product of connected correlators with the right large $N$ scaling. Moreover, as we did for the free energy, it proves convenient to subtract the result from the Gaussian matrix model, which in this context corresponds to the ${\cal N}=4$ theory.

\begin{figure}[ht]
\begin{minipage}[t]{.4\textwidth}
\vspace{-7ex}\begin{subfigure}{0.9\linewidth}
    \centering
    \includegraphics[scale=.02]{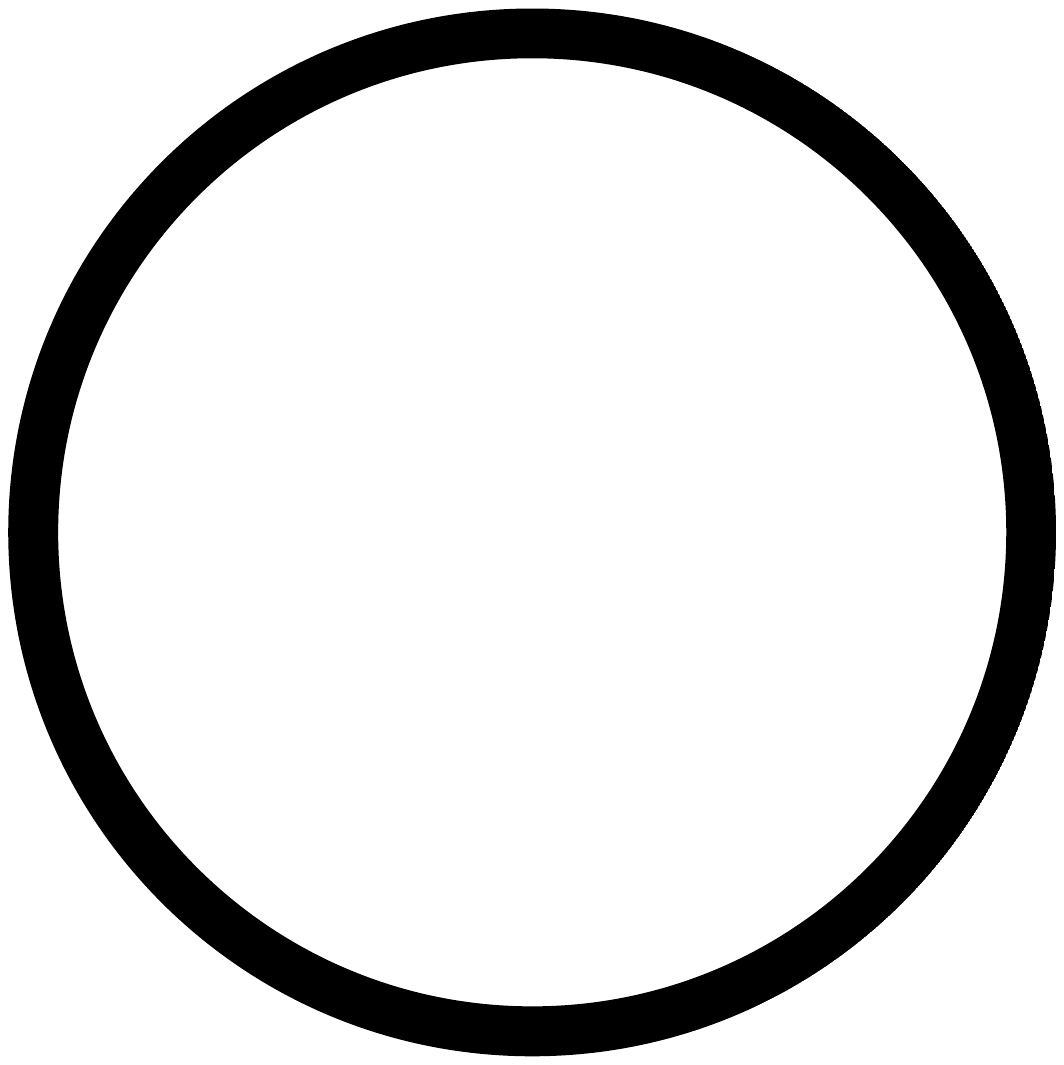}
  \end{subfigure}
  
\vspace{-3ex}\begin{subfigure}{0.9\linewidth}
   \centering
    \includegraphics[scale=.1]{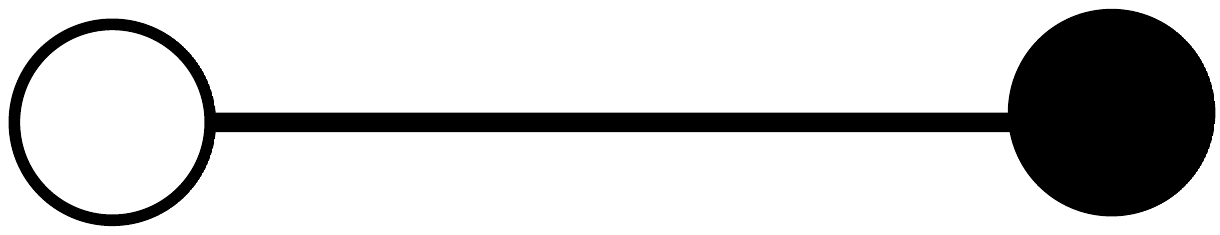}
  \end{subfigure}
  
\vspace{-15ex}\begin{subfigure}{0.9\linewidth}
   \centering
    \includegraphics[scale=.18]{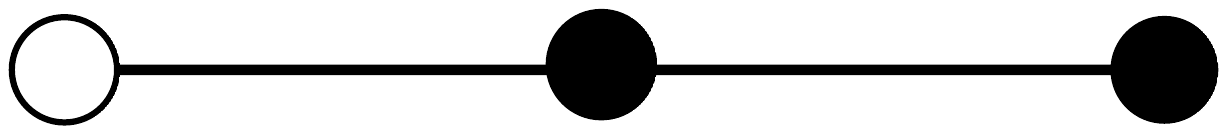}
    \end{subfigure}
    
\vspace{-21ex}\begin{subfigure}{0.9\linewidth}
   \centering
    \includegraphics[scale=.18]{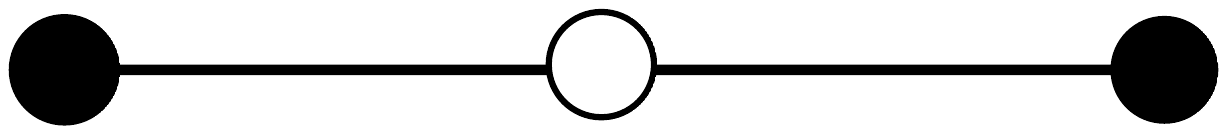}
  \end{subfigure}   
\end{minipage}%
\begin{minipage}[t]{.3\textwidth}
\begin{subfigure}{.2\linewidth}
\centering
    \includegraphics[scale=.11]{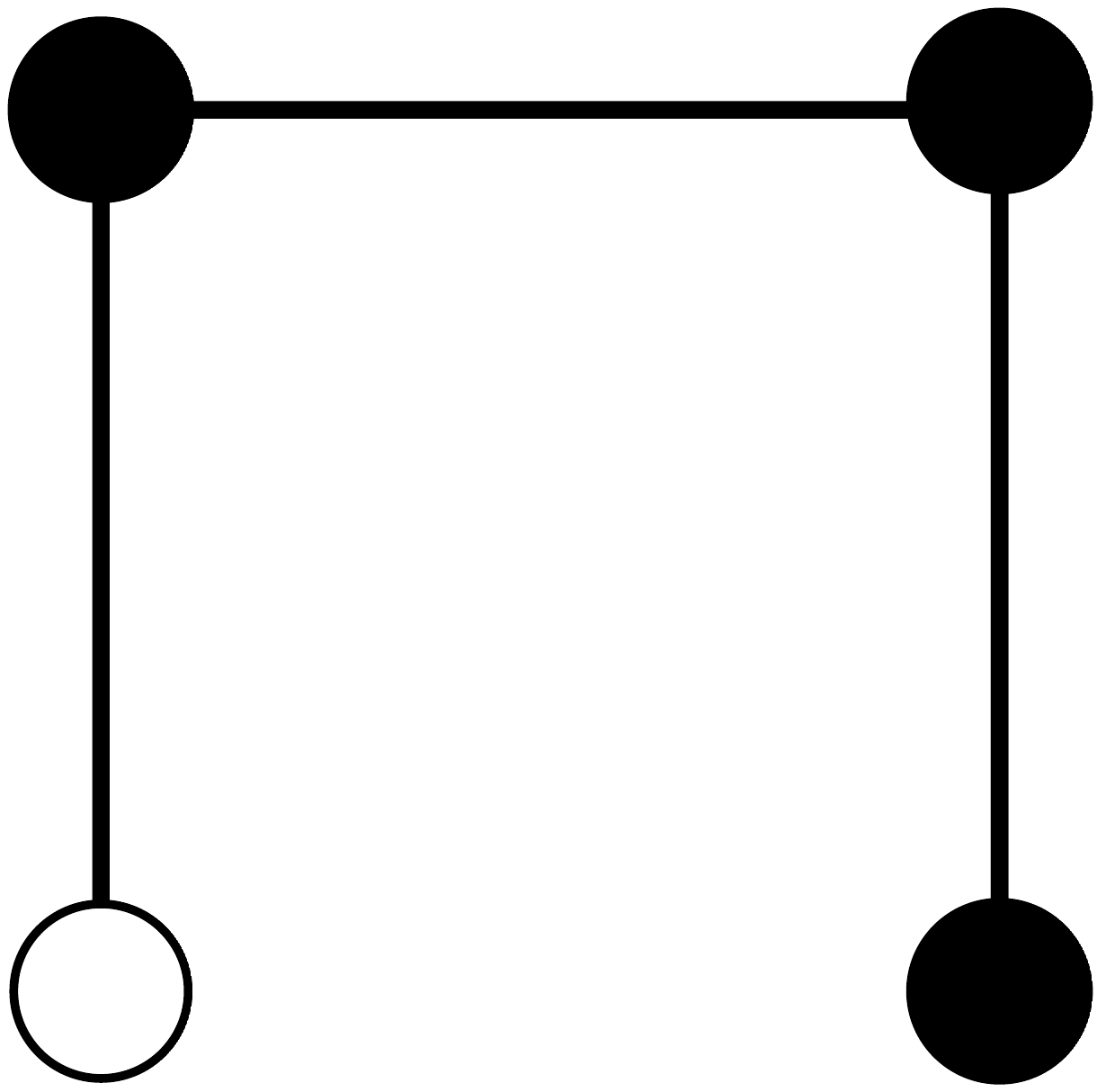}    
  	  \end{subfigure} 
  	  
\begin{subfigure}{.2\linewidth}
\centering
    \includegraphics[scale=.11]{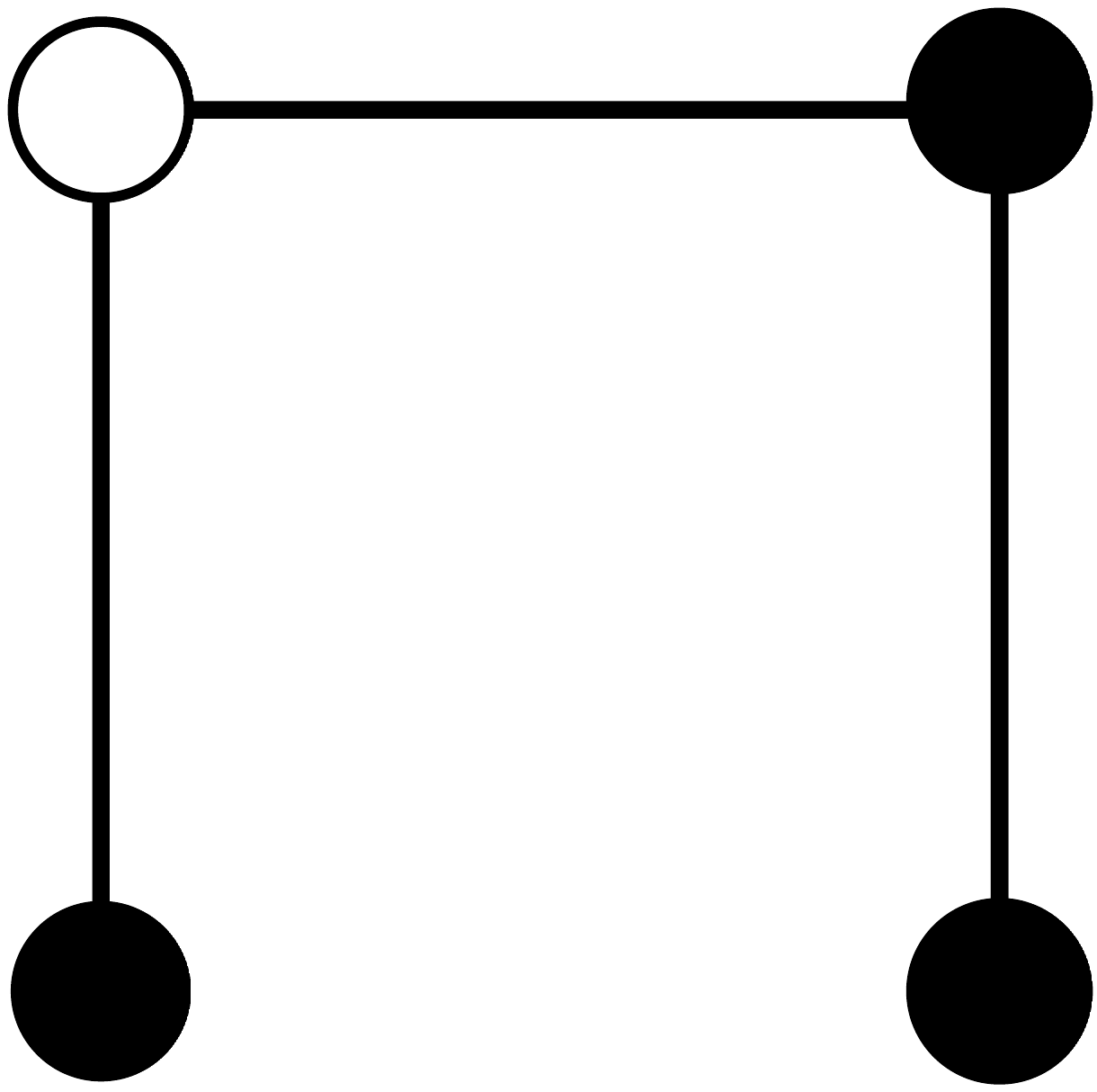}    
  	  \end{subfigure}  
\end{minipage}%
\begin{minipage}[t]{.3\textwidth} 
\begin{subfigure}{.2\linewidth}
\centering
    \includegraphics[scale=.11]{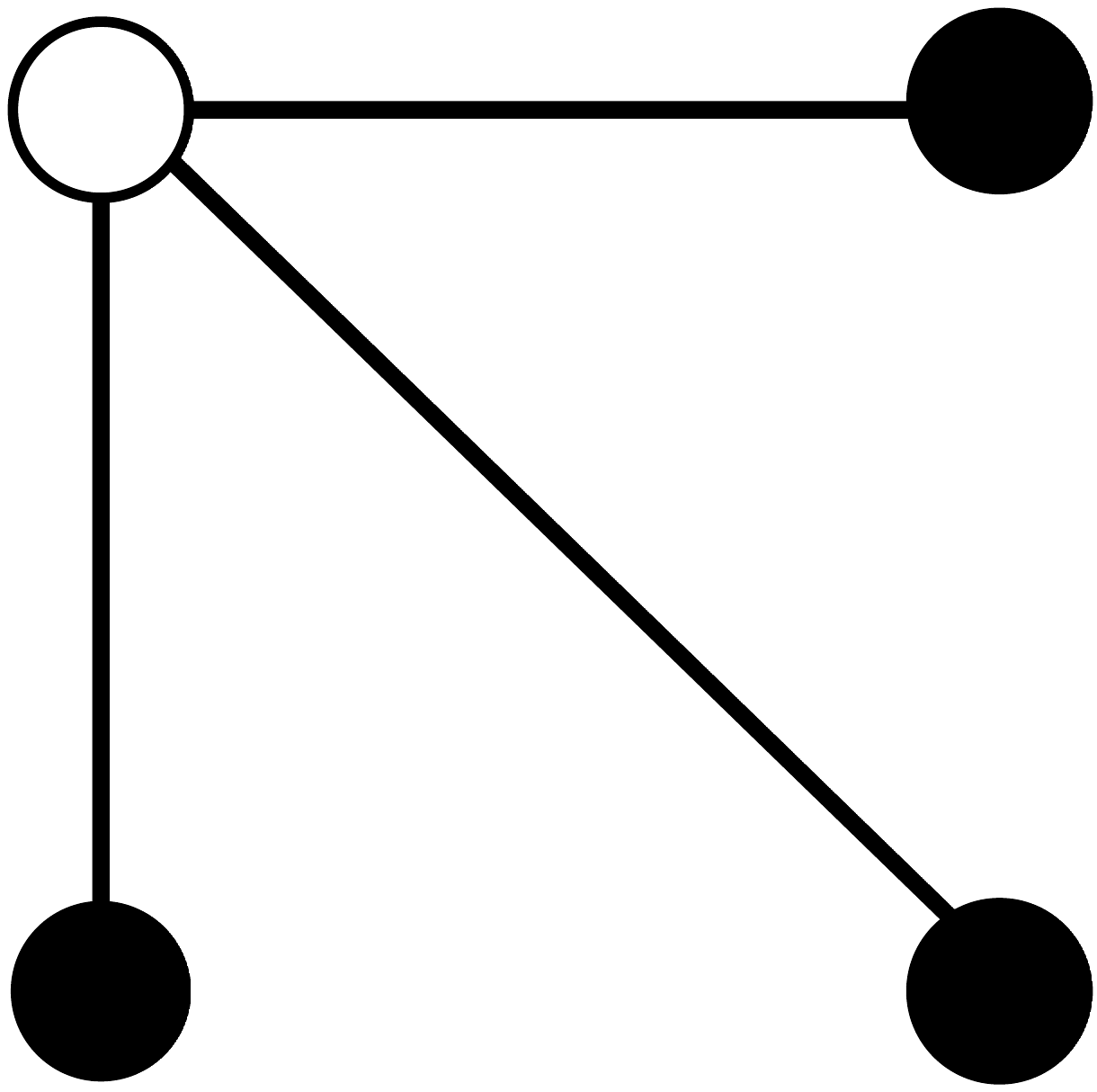}    
  	  \end{subfigure}  
  	  
\hspace{-4ex}\begin{subfigure}{.2\linewidth}
\centering
    \vspace{-5ex}\includegraphics[scale=.18]{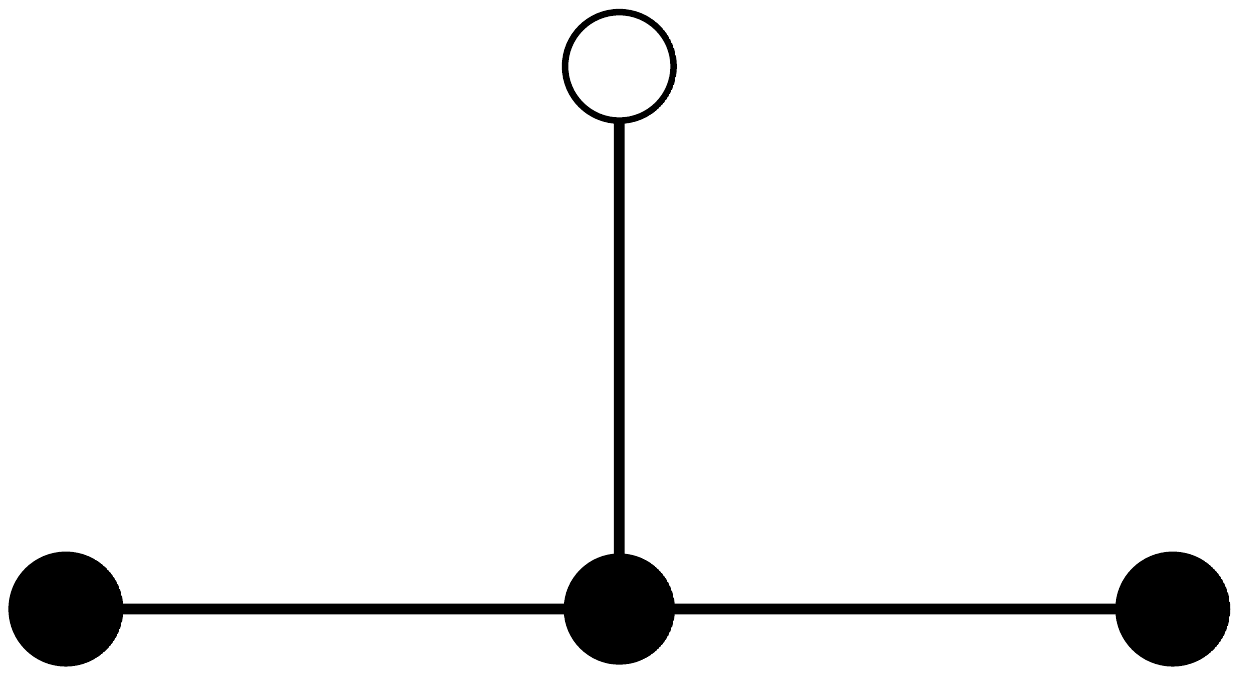}    
  \end{subfigure}
\end{minipage}
\vspace{-8ex}\caption{The list of rooted trees up to 4 vertices. In each tree, the distinguished vertex is represented by the white dot.}
\label{rootedtrees}
\end{figure}

After expanding the effective action, we want to extract the piece of $\vev{\traza{2l}S^m}$ that scales like $N$. This correlator contains $2m+1$ traces, so by the same argument as in the previous section, a product of $s$ connected correlators scales like $2s-2m-1$, which implies the relevant piece are again products of $m+1$ connected correlators. On the other hand, $\traza{2l}$ can not be by itself in one of these correlators, because it will be cancelled by the ${\cal N}=4$ subtraction. As a consequence, after fixing the correlator that contains $\traza{2l}$, we are again distributing $2m$ operators into $m+1$ connected correlators, so again we find a tree structure! There is however, an important difference: now, one of the connected correlators contains $\traza{2l}$, so it is distinguished from the rest. When we translate the product of connected correlators to a tree graph, we have to distinguish one of the vertices, the one that correspond to the correlator containing $\traza{2l}$. In the mathematical literature, trees with a distinguished vertex are call rooted trees \cite{harary}. See figure (\ref{rootedtrees}) for the list of rooted trees with up to four vertices. The number of rooted trees with $n$ vertices is \cite{rootedtrees}
\be 
1,1,2,4,9,20,48,\dots
\ee

All in all, we find

\begin{multline}
\vev{W}_{\mathcal{N}=2}-\vev{W}_{\mathcal{N}=4} = \sum_{l=1}^{\infty}\frac{b^{2l}}{(2l)!} \left(\frac{\lambda}{4}\right)^l
\sum_{m=1}^\infty 
\frac{(-\beta_G)^m}{m!}\sum_{n_1,\dots,n_m=2}^\infty \left( \frac{-\lambda}{16 \pi^2} \right)^{n_1+\dots+n_m}\frac{\zeta(2n_1-1)\dots \zeta(2n_m-1)}{n_1\dots n_m} \\
\sum_{k_1=1}^{n_1-1} {2n_1\choose 2k_1} \dots \sum_{k_m=1}^{n_m-1} {2n_m \choose 2k_m}  
\sum_{\substack{\text{directed rooted trees}\\ \text{with m labeled edges}}}
\prod_{i=1}^{m+1} {\cal V}_i
\end{multline}
By the same arguments that we used in the discussion of the free energy, this expression can be simplified to a sum over unlabeled rooted trees
\begin{multline}
\vev{W}_{\mathcal{N}=2}-\vev{W}_{\mathcal{N}=4} = \sum_{l=1}^{\infty}\frac{b^{2l}}{(2l)!} \left(\frac{\lambda}{4}\right)^l
\sum_{m=1}^\infty 
(- 2\beta_G)^m \sum_{n_1,\dots,n_m=2}^\infty \left( \frac{-\lambda}{16 \pi^2} \right)^{n_1+\dots+n_m}\frac{\zeta(2n_1-1)\dots \zeta(2n_m-1)}{n_1\dots n_m} \\
\sum_{k_1=1}^{n_1-1} {2n_1\choose 2k_1} \dots \sum_{k_m=1}^{n_m-1} {2n_m \choose 2k_m}  
\sum_{\substack{\text{unlabeled rooted}\\ \text{trees with m edges}}}
\frac{1}{|\text{Aut(T)}|} \prod_{i=1}^{m+1} {\cal V}_i
\label{finalplanarw}
\end{multline}

To illustrate (\ref{finalplanarw}), let's work out the terms up to $\beta_G^3$,
%\begin{multline}
%\vev{W}_{\mathcal{N}=2}-\vev{W}_{\mathcal{N}=4} =\sum_{l=1}^\infty \frac{b^{2l}}{l!(l-1)!} \left(\frac{\lambda}{4}\right)^l
%\left[- \beta \sum_{n=2}^\infty \left(\frac{-\lambda}{16 \pi^2}\right)^n
%\frac{\zeta(2n-1)}{n} {2n \choose n} \sum_{k=1}^{n-1} {n \choose k}{n \choose k-1} \frac{2}{l+k} \right. \\
%+ \beta^2\sum_{n_1,n_2=2}^\infty \left(\frac{-\lambda}{16 \pi^2}\right)^{n_1+n_2} \frac{\zeta(2n_1-1) \zeta(2n_2-1)}{n_1 n_2} {2n_1 \choose n_1} {2n_2 \choose n_2} 
%\sum_{k=1}^{n_1-1} {n_1 \choose k_1}{n_1 \choose k_1-1} \sum_{k=2}^{n_2-1} {n_2 \choose k_2}{n_2 \choose k_2-1} \\
% \left(4 \frac{(n_1-k_1+1)(n_1-k_1)}{(l+n_1-k_1)(k_1+k_2)}+2\right)-8\beta^ 3 \sum_{n_1,n_2,n_3=2}^\infty \left(\frac{-\lambda}{16 \pi^2}\right)^{n_1+n_2+n_3} \frac{\zeta(2n_1-1) \zeta(2n_2-1)\zeta(2n_3-1)}{n_1 n_2 n_3} \\
%\frac{(2n_1)!}{(n_1-1)!^2}\frac{(2n_2)!}{(n_2-1)!^2}\frac{(2n_3)!}{(n_3-1)!^2}
%\sum_{k=1}^{n_1-1} {n_1-1 \choose k_1}{n_1-1 \choose k_1-1} \sum_{k=2}^{n_2-1} {n_2-1 \choose k_2}{n_2-1 \choose k_2-1} \sum_{k=3}^{n_3-1} {n_3-1 \choose k_3}{n_3-1 \choose k_3-1} \\
%\left( \frac{1}{(l+k_1)(n_1-k_1+k_2)(n_2-k_2+k_3)(n_3-k_3+1)(n_3-k_3)}+\frac{1}{2}\frac{1}{k_2 k_3 (l+k_1)(k_2+1)(k_3+1)}\right.\\
%\left. \left. +\frac{1}{3!}\frac{l+n_1-k_1+n_2-k_2+n_3-k_3-1}{k_1 k_2 k_3 (k_1+1)(k_2+1)(k_3+1)}+\frac{1}{k_1(k_1+1)(n_2-k_2+k_3)(n_3-k_3+1)(n_3-k_3)}\right) +\cdots\right]
%\end{multline}
\begin{multline}
\vev{W}_{\mathcal{N}=2}-\vev{W}_{\mathcal{N}=4} =\sum_{l=1}^\infty \frac{b^{2l}}{l!(l-1)!} \left(\frac{\lambda}{4}\right)^l
\left[- \beta_G \sum_{n=2}^\infty \left(\frac{-\lambda}{16 \pi^2}\right)^n
\frac{\zeta(2n-1)}{n} {2n \choose n} \sum_{k=1}^{n-1} {n \choose k}{n \choose k-1} \frac{2}{l+k} \right. \\
+ \beta_G^2\sum_{n_1,n_2=2}^\infty \left(\frac{-\lambda}{16 \pi^2}\right)^{n_1+n_2} \frac{\zeta(2n_1-1) \zeta(2n_2-1)}{n_1 n_2} {2n_1 \choose n_1} {2n_2 \choose n_2} 
\sum_{k=1}^{n_1-1} {n_1 \choose k_1}{n_1 \choose k_1-1} \sum_{k=2}^{n_2-1} {n_2 \choose k_2}{n_2 \choose k_2-1} \\
 \left(4 \frac{(n_1-k_1+1)(n_1-k_1)}{(l+n_1-k_1)(k_1+k_2)} + 2 \right) - \frac{\beta_G^3}{6} \sum_{n_1,n_2,n_3=2}^\infty \left(\frac{-\lambda}{16 \pi^2}\right)^{n_1+n_2+n_3} \frac{\zeta(2n_1-1) \zeta(2n_2-1)\zeta(2n_3-1)}{n_1 n_2 n_3} 
 \\
{2n_1 \choose n_1}{2n_2 \choose n_2}{2n_3 \choose n_3} \sum_{k=1}^{n_1-1} {n_1 \choose k_1}{n_1 \choose k_1-1} \sum_{k=2}^{n_2-1} {n_2 \choose k_2}{n_2 \choose k_2-1} \sum_{k=3}^{n_3-1} {n_3 \choose k_3}{n_3 \choose k_3-1} \\
\left( 48\frac{(n_1-k_1)(n_1-k_1+1)(n_3-k_3)(n_3-k_3+1)}{(l+k_1)(k_2+k_3)(n_1-k_1+n_3-k_3)} + 48\frac{(n_1-k_1)(n_1-k_1+1)}{k_3+n_1-k_1} \right. \\
\left. \left. + 24\frac{(n_1-k_1)(n_1-k_1+1)}{l+n_1-k_1} + 8(l+k_1+k_2+k_3-1)\right) +\cdots \right]
\end{multline}
where the dots stand for terms with more than three values of the $\zeta$ function. We have checked that this expression correctly reproduces the explicit results of Appendix A of \cite{Passerini:2011fe}, where this expectation value was computed up to $\lambda^7$. This match provides a very non-trivial check of our computation.

Finally, let's consider an example of a theory with $\beta_G=0$, namely the SU($N$) gauge theory with a $2-$symmetric and a $2-$anti-symmetric hypermultiplet. This theory has $\alpha_G = \beta_G = 0$ and $\gamma_G = -2$, so in the planar limit $\langle W \rangle_{\mathcal{N}=2} = \langle W \rangle_{\mathcal{N}=4}$, and we can compute some subleading $1/N^2$ terms in $\langle W \rangle_{\mathcal{N}=2}$. In particular, we  will now derive the term linear in $\gamma_G$, so it contains all the terms with a single value of the $\zeta$ function. To do so note that in this case the effective action (\ref{def:general_action}) has only odd powers so we need the equivalent result of (\ref{gopakumarf}) with two odd powers and a single even power. This is \cite{Gopakumar:2012ny}
\be 
\vev{\traza{2k_1+1}\traza{2k_2+1}\traza{2k}}_c = \frac{(2k_1+1)!}{k_1!^2} \frac{(2k_2+1)!}{k_2!^2} \frac{(2k_3)!}{(k_3-1)! k_3!}
\ee

With this in mind we obtain
	\begin{equation}
		\langle W \rangle_{\mathcal{N}=2} - \langle W \rangle_{\mathcal{N}=4} = - \frac{\gamma_G}{N^2} \sum_{l=1}^\infty \frac{(4\pi^2 b^2)^l}{l!(l-1)!} \sum_{n=3}^\infty \zeta(2n-1)(-1)^n {2n \choose n-1}(n+1)\left(\frac{\lambda}{16\pi^2} \right)^{n+l} \sum_{k=1}^{n-2} {n-1 \choose k}^2 \,,
	\end{equation}
where in fact the sums over $k$ and $l$ can be performed, which gives us
\be
\langle W \rangle_{\mathcal{N}=2} - \langle W \rangle_{\mathcal{N}=4} = - \frac{\gamma_G}{N^2} \frac{b\sqrt{\lambda}}{2}I_1(b\sqrt{\lambda}) \sum_{n=3}^\infty \zeta(2n-1)(-1)^n {2n \choose n-1}(n+1)\left[ {2(n-1) \choose n-1} - 2 \right]\left( \frac{\lambda}{16\pi^2} \right)^n \,.
\ee
Where $I_1(x)$ is the modified Bessel function of the first kind and this result is computed for SU($N$). In the large $N$ limit, $n$-point functions of traces of odd powers don't coincide for U($N$) and SU($N$), so this result can't be compared with the large $N$ limit of (\ref{res:gg6}).

\acknowledgments
We would like to thank  Igor Klebanov, Marcos Mari\~no, Jorge Russo, Mikel S\'anchez, Albert Sanglas and Nikolaos Triantafyllou for useful discussions on related topics. We are specially thankful to Marcos Mari\~no for sharing with us a Mathematica notebook implementing the computations of Appendix B of \cite{Grassi:2014vwa}, which allowed us to check our result (\ref{finalfree}) up to $\lambda^7$. Research supported by Spanish MINECO under projects MDM-2014-0369 of ICCUB (Unidad de Excelencia ``Mar\'ia de Maeztu") and FPA2017-76005-C2-P, and by AGAUR, grant 2017-SGR 754.  J. M. M. is further supported by "la Caixa" Foundation (ID 100010434) with fellowship code LCF/BQ/IN17/11620067, and from the European Union's Horizon 2020 research and innovation programme under the Marie Sk{\l}odowska-Curie grant agreement No. 713673. A. R. F. is further supported by an FPI-MINECO fellowship. 

\appendix

\section{$\mathcal{Z}_{\text{1-loop}}$ and $S_{int}^G$ for the classical groups}
In this appendix we present the full expression of the $\mathcal{Z}_{\text{1-loop}}$ for (\ref{def:1_loop}) and the corresponding expression of the effective action
\be
\begin{split}
Z_{\text{1-loop}}^{\mathrm{SU}(N)}&=
\frac{\prod_{u<v=1}^N H(ia_u-ia_v)^2}
{\prod_{u<v=1}^N H(ia_u-ia_v)^{2 n_{\mathrm{adj}}} \prod_{u=1}^N H(ia_u)^{n_f} \prod_{u\leq v=1}^N H(ia_u+ia_v)^{n_{\mathrm{sym}}} \prod_{u<v=1}^N H(ia_u+ia_v)^{n_{\mathrm{asym}}}} \\
\label{def:1_loop_sun}
\end{split}
\ee
\be
Z_{\text{1-loop}}^{\mathrm{SO}(2N)} = \frac{\prod_{u<v}^N H^2 (i a_u + i a_v) H^2 (i a_u - i a_v) }{\prod_{u<v}^N H(i a_u + i a_v)^{2n_{\mathrm{adj}}}  H(i a_u - i a_v)^{2n_{\mathrm{adj}}}\prod_{u=1}^N H(i a_u)^{2n_{\mathrm{v}}}}
\label{def:1_loop_so2n}
\ee
\be
Z_{\text{1-loop}}^{\mathrm{SO}(2N+1)} = \frac{\prod_{u<v}^N H^2 (i a_u + i a_v) H^2 (i a_u - i a_v) \prod_{u=1}^N H(i a_u)^2 }{\prod_{u<v}^N H(i a_u + i a_v)^{2n_{\mathrm{adj}}}  H(i a_u -ia_v)^{2n_{\mathrm{adj}}}\prod_{u=1}^N H(i a_u)^{2n_{\mathrm{adj}}+2n_{\mathrm{v}}}} 
\label{def:1_loop_so2n1}
\ee
\be
Z_{\text{1-loop}}^{\mathrm{Sp}(N)}=\frac{\prod_{u<v}^N H^2 (i a_u + i a_v) H^2 (i a_u - i a_v) \prod_{u=1}^N H(2i a_u)^2 } {\prod_{u<v}^N H(i a_u + i a_v)^{2n_{\mathrm{adj}}+2n_a}  H(i a_u -ia_v)^{2n_{\mathrm{adj}}+2n_a} \prod_{u=1}^N H(2i a_u)^{2n_{\mathrm{adj}}}  \prod_{u=1}^N H(i a_u)^{2n_{\mathrm{v}}}},
\label{def:1_loop_spn}
\ee

\begin{multline}
S^{\mathrm{SU}(N)}= \sum_{n=2}^\infty \frac{\zeta(2n-1)(-1)^n}{n}
\left\{\frac{4-4^n}{2}(n_{\mathrm{sym}}-n_{\mathrm{asym}}) \traza{2n} \right. \\
\left. + \sum_{k=2}^{2n-2} {2n \choose k} \left((-1)^k (1-n_{\mathrm{adj}})-\frac{n_{\mathrm{sym}}+n_{\mathrm{asym}}}{2}\right) \traza{2n-k} \traza{k} \right\},
\label{eq:interaction_action}
\end{multline}

\begin{equation}
	S^{\mathrm{SO}(2N)} =  \sum_{n=2}^\infty \frac{\zeta(2n-1)(-1)^n}{n} (1-n_{\mathrm{adj}}) \left\{ \sum_{k=2}^{2n-2} {2n \choose k} \traza{2n-k} \traza{k} \left(1+(-1)^k \right) + (4-4^n) \traza{2n} \right\}
\end{equation}

\begin{equation}
	S^{\mathrm{SO}(2N+1)} =  \sum_{n=2}^\infty \frac{\zeta(2n-1)(-1)^n}{n} (1-n_{\mathrm{adj}}) \left\{ \sum_{k=2}^{2n-2} {2n \choose k} \traza{2n-k} \traza{k} \left(1+(-1)^k \right) + (4-4^n) \traza{2n} \right\}
\end{equation}

\begin{multline}
	S^{\mathrm{Sp}(2N)} =  \sum_{n=2}^\infty \frac{\zeta(2n-1)(-1)^n}{n}  \left\{ (1-n_{\mathrm{adj}}-n_{\mathrm{asym}})\sum_{k=2}^{2n-2} {2n \choose k} \traza{2n-k} \traza{k} \left(1+(-1)^k \right) \right. \\
	\left. + (4-4^n) (n_{\mathrm{adj}}-n_{\mathrm{asym}}-1) \traza{2n} \right\}
\end{multline}

\section{Explicit planar free energy up to 13th order}

In this appendix we present the result of evaluating (\ref{finalfree}) up to $\lambda^{13}$. To do so, we use the shorthand $\tlambda =-\frac{\lambda}{16 \pi^2}$. Furthermore, we are not writing powers of $\beta_G$; to recover them, write one $\beta_G$ for each $\zeta$. The planar free energy is then
\begin{multline*}
F_0(\lambda)-F_0(\lambda)^{{\cal N}=4}=-3 \zeta_3 \tlambda^2 - 20 \zeta_5 \tlambda^3 +(36 \zeta_3^2-140 \zeta_7)\tlambda^4+
(720 \zeta_3 \zeta_5-1092 \zeta_9)\tlambda^5 \\
+(-720 \zeta_3^3+3800 \zeta_5^2+6720 \zeta_3 \zeta_7-9394 \zeta_{11})\tlambda^6+(-25920 \zeta_3^2 \zeta_5+73360 \zeta_5 \zeta_7+65520 \zeta_3 \zeta_9-87516 \zeta_{13})\tlambda^7 \\
+(18144 \zeta_3^4-316800 \zeta_3 \zeta_5^2-282240 \zeta_3^2 \zeta_7+361620 \zeta_7^2+732480 \zeta_5 \zeta_9+676368 \zeta_3 \zeta_{11}-868725 \zeta_{15})\tlambda^8 \\
+(967680 \zeta_3^3 \zeta_5-(3920000 \zeta_5^3)/3-6968640 \zeta_3 \zeta_5 \zeta_7-3144960 \zeta_3^2 \zeta_9+7331520 \zeta_7 \zeta_9+7700880 \zeta_5 \zeta_{11}+\\
7351344 \zeta_3 \zeta_{13}-9072492 \zeta_{17})\tlambda^9 +(-(2612736/5) \zeta_3^5+19440000 \zeta_3^2 \zeta_5^2+11612160 \zeta_3^3 \zeta_7-43394400 \zeta_5^2 \zeta_7 \\
-38478720 \zeta_3 \zeta_7^2-78180480 \zeta_3 \zeta_5 \zeta_9+37570176 \zeta_9^2-36523872 \zeta_3^2 \zeta_{11}+77994840 \zeta_7 \zeta_{11}\\
+84942000 \zeta_{5} \zeta_{13}+83397600 \zeta_{3} \zeta_{15}-(493668032 \zeta_{19})/5)\tlambda^{10} \\
+(-37324800 \zeta_3^4 \zeta_5+173952000 \zeta_3 \zeta_5^3+466502400 \zeta_3^2 \zeta_5 \zeta_7-481376000 \zeta_5 \zeta_7^2+141523200 \zeta_3^3 \zeta_9-489014400 \zeta_5^2 \zeta_9\\
-865186560 \zeta_3 \zeta_7 \zeta_9-912859200 \zeta_3 \zeta_5 \zeta_{11}+806319360 \zeta_9 \zeta_{11}-441080640 \zeta_3^2 \zeta_{13}+868659792 \zeta_7 \zeta_{13}\\
+975477360 \zeta_5 \zeta_{15} +979829136 \zeta_3 \zeta_{17}-1111643260 \zeta_{21})\tlambda^{11} \\
+(16422912 \zeta_3^6-1064448000 \zeta_3^3 \zeta_5^2+584160000 \zeta_5^4-479001600 \zeta_3^4 \zeta_7+6253228800 \zeta_3 \zeta_5^2 \zeta_7+2787966720 \zeta_3^2 \zeta_7^2 \\
-(5345751040 \zeta_7^3)/3 +5678830080 \zeta_3^2 \zeta_5 \zeta_9-10857759360 \zeta_5 \zeta_7 \zeta_9-4866160320 \zeta_3 \zeta_9^2+1785611520 \zeta_3^3 \zeta_{11} \\
-5731228800 \zeta_5^2 \zeta_{11} -10116912960 \zeta_3 \zeta_7 \zeta_{11}+4356229416 \zeta_{11}^2-11075201280 \zeta_3 \zeta_5 \zeta_{13}+ 9045036000 \zeta_9 \zeta_{13}\\
-5504241600 \zeta_3^2 \zeta_{15}+10057407360 \zeta_7 \zeta_{15}+11579728160 \zeta_5 \zeta_{17}+11848032768 \zeta_3 \zeta_{19}-(38632924694 \zeta_{23})/3)\tlambda^{12} \\
+(1478062080 \zeta_3^5 \zeta_5-15137280000 \zeta_3^2 \zeta_5^3-27186001920 \zeta_3^3 \zeta_5 \zeta_7+27942656000 \zeta_5^3 \zeta_7+74609203200 \zeta_3 \zeta_5 \zeta_7^2 \\
-6227020800 \zeta_3^4 \zeta_9+75975782400 \zeta_3 \zeta_5^2 \zeta_9+67576965120 \zeta_3^2 \zeta_7 \zeta_9-60299164800 \zeta_7^2 \zeta_9-61206969600 \zeta_5 \zeta_9^2 \\
+71569681920 \zeta_3^2 \zeta_5 \zeta_{11}-127317072960 \zeta_5 \zeta_7 \zeta_{11}-113820094080 \zeta_3 \zeta_9 \zeta_{11}+23289057792 \zeta_3^3 \zeta_{13}\\
-69769814400 \zeta_5^2 \zeta_{13}-122896597824 \zeta_3 \zeta_7 \zeta_{13}+98300538336 \zeta_{11} \zeta_{13}-138770723520 \zeta_3 \zeta_5 \zeta_{15}+ \\
105367232880 \zeta_9 \zeta_{15} -70547697792 \zeta_3^2 \zeta_{17}+120227642080 \zeta_7 \zeta_{17}+141264773520 \zeta_5 \zeta_{19}
+146736910320 \zeta_{3} \zeta_{21} \\
-152833845400 \zeta_{25})\tlambda^{13} + {\cal O}(\tlambda^{14})
\end{multline*}

Using the method of orthogonal polynomials explained in Appendix B of \cite{Grassi:2014vwa} , we have checked this result up to $\tlambda^7$.

\bibliographystyle{JHEP}

\end{document}